# ESTIMATION OF NANOMETER-THICKNESS LAYER OF 6B GRAPHITE: AN EXPERIMENTAL ACTIVITY TO STUDY OHM'S LAW IN HIGHER EDUCATION

Paulo Henrique Eleuterio Falsetti[1], André Coelho da Silva[2], Leonardo Geraldino da Silva[3], Douglas Mendes da Silva Del Duque[2], Murilo Antonio Menegati[3], Bruno Fernando Gianelli[2], Vagner Romito de Mendonça[2], Idelma Aparecida Alves Terra[4]

[1] Federal University of São Carlos (UFSCar) – Sorocaba Campus – SP
[2] Federal Institute of Education, Science and Technology of São Paulo (IFSP) – Itapetininga Campus – SP
[3] Federal Institute of Education, Science and Technology of São Paulo (IFSP) – Piracicaba Campus – SP
[4] National Service for Commercial Learning (Senac) – Piracicaba – SP

# ABSTRACT

Ohm's Law is crucial for understanding electrical circuits and conductive materials. Formulated by Georg Simon Ohm in the 19th century, it describes the direct proportional relationship between electric voltage and electric current in ohmic conductors. Despite its apparent simplicity, the literature has pointed out that students at different educational levels have difficulty understanding it, especially due to the abstraction associated with the concepts of electric voltage, electric current, and electrical resistance. In order to provide possibilities to overcome this abstraction and the associated learning difficulties, the present work proposes an experimental activity that applies Ohm's Law to determine the nanometer-scale thickness of 6B graphite traces deposited on tracing paper. The employed methodology is based on measuring physical quantities such as electric voltage, electric current, and length, and on analyzing the collected data. The thickness of the graphite traces determined by this method was also compared with values obtained using a Scanning Probe Microscope (SPM). Based on the proposed methodology, the thickness of the traces was determined to be 456.5(34) nm, while SPM measurements yielded an average thickness of 440(50) nm. Thus, the results show good agreement between the measurements within experimental uncertainties, validating the effectiveness of the suggested methodology and indicating that it is a viable proposal for Electricity courses in Higher Education and, with adaptations, even for High School.

# ESTIMATIVA DE ESPESSURA NANOMÉTRICA DE CAMADA DE GRAFITE 6B: UMA ATIVIDADE EXPERIMENTAL PARA ESTUDAR AS LEI DE OHM NO ENSINO SUPERIOR

Paulo Henrique Eleuterio Falsetti[1], André Coelho da Silva[2], Leonardo Geraldino Da Silva[3], Douglas Mendes da Silva Del Duque[2], Murilo Antonio Menegati[3], Bruno Fernando Gianelli[2], Vagner Romito de Mendonça[2], Idelma Aparecida Alves Terra[4]

[1] Universidade Federal de São Carlos (UFSCar) – Campus Sorocaba – SP

[2] Instituto Federal de Educação, Ciência e Tecnologia de São Paulo (IFSP) – Campus Itapetininga – SP

[3] Instituto Federal de Educação, Ciência e Tecnologia de São Paulo (IFSP) – Campus Piracicaba – SP

[4] Serviço Nacional de Aprendizagem Comercial (Senac) – Piracicaba – SP

# RESUMO

A Lei de Ohm é crucial para a compreensão de circuitos elétricos e materiais condutores. Formulada por Georg Simon Ohm no século XIX, ela descreve a relação de proporcionalidade direta entre tensão elétrica e corrente elétrica em condutores ôhmicos. Apesar de sua aparente simplicidade, a literatura tem apontado que estudantes de diferentes níveis de ensino possuem dificuldade em compreendê-la, especialmente por conta da abstração associada aos conceitos de tensão elétrica, corrente elétrica e resistência elétrica. Com o intuito de oferecer possibilidades para superar essa abstração e as dificuldades de aprendizagem, o presente trabalho propõe uma atividade experimental que utiliza a Lei de Ohm para determinar a espessura nanométrica de traços de grafite do tipo 6B depositados sobre papel vegetal. A metodologia empregada se baseia na medição de grandezas físicas como tensão elétrica, corrente elétrica e comprimento e na análise dos dados coletados. A espessura dos traços de grafite determinada dessa maneira foi também comparada com valores obtidos utilizando um microscópio de varredura por sonda (SPM). Com base na metodologia proposta, determinou-se a espessura dos traços como sendo de 456,5(34) nm. Já com o uso do SPM, a espessura média dos mesmos traços foi medida como sendo de 440(50) nm. Dessa forma, os resultados evidenciam que houve uma boa concordância entre as medições dentro das incertezas experimentais, validando a eficácia da metodologia sugerida e indicando que se trata de uma proposta viável para disciplinas de Eletricidade no Ensino Superior e, com adaptações, até mesmo para o Ensino Médio.

**Palavras-chaves:** Lei de Ohm; Eletricidade; Resistividade; Prática experimental; Ensino Superior.

# INTRODUÇÃO

Há certo consenso entre professores e pesquisadores da área de Ensino de Física sobre a relevância da experimentação enquanto abordagem didática (1). Estudos como o de Parreira e Dickman (2020) evidenciam que tanto docentes quanto discentes dos cursos de Engenharia reconhecem, de forma unânime, a importância dos experimentos de Física para estimular uma maior participação dos estudantes em atividades acadêmicas, além de contribuir significativamente para a consolidação dos conceitos teóricos por meio de sua aplicação prática (2). Ademais, os estudantes que participaram da pesquisa conduzida pelos referidos autores também apontaram o auxílio das práticas experimentais no que tange ao relacionamento das teorias estudadas com o cotidiano (2).

De fato, a utilização de experimentos didáticos no Ensino de Ciências desempenha um papel fundamental na aprendizagem dos alunos. Através da prática, estimulados pela curiosidade, eles podem explorar novos conhecimentos, levantar questionamentos sobre diferentes temas e desenvolver uma aprendizagem mais significativa (3). A prática experimental lhes permite também conectar a teoria ao cotidiano de forma prática, incentivando a reflexão e a expressão de ideias (3). Considerando os conhecimentos prévios dos alunos, as práticas experimentais podem contribuir para a assimilação de novos conhecimentos, levando-se em consideração que a praticidade da atividade pode ser um atrativo (3). Esse tipo de experimentação se relaciona à Teoria da Aprendizagem Significativa proposta por David Paul Ausubel, segundo a qual o aluno atribui sentido pessoal ao que aprende ao relacionar as novas informações ao seu conhecimento prévio. Quando essa conexão não ocorre, o aprendizado se torna mecânico, sem integração real com a estrutura cognitiva. Nesses casos, o conhecimento é memorizado sem compreensão profunda (4).

As atividades experimentais tendem a estimular a participação ativa, despertando curiosidade e tornando a aprendizagem mais envolvente e motivadora (5). Assim, torna-se uma estratégia eficaz para superar dificuldades, favorecendo a construção de conhecimentos e o desenvolvimento de habilidades científicas (5). Além do mais, conforme sinalizado por Carrascosa (2006) (6), é relevante pensar em atividades experimentais que possuam teor investigativo, com roteiros não muito fechados, os quais poderiam limitar a reflexão e a criatividade, como se fossem simples receituários a se seguir mecanicamente.

A Lei de Ohm se apresenta como uma relação matemática muito abstrata para os alunos, sendo necessárias diferentes abordagens para a consolidação do seu significado. Na literatura, pode-se encontrar diversos trabalhos que utilizam a prática experimental para abordar a Lei de Ohm. O estudo de Prastyaningrum & Pratama (2019) (7) investigou as concepções dos alunos sobre a Lei de Ohm utilizando um método qualitativo, com coleta de dados por meio de perguntas dissertativas e entrevistas. A análise dos resultados mostrou que os estudantes de engenharia elétrica não compreendem totalmente o conceito da Lei de Ohm, principalmente devido à dificuldade em interpretar a linguagem da física aplicada aos fenômenos da eletricidade. O estudo também destacou a importância do ensino baseado em projetos como estratégia para reduzir concepções errôneas e aprimorar a compreensão dos fundamentos da eletricidade.

Outros estudos – como Rocha & Santiago (2017) (8) e Oliveira (2015) (9) - abordam a Lei de Ohm determinando a resistência e/ou a resistividade de materiais com base na análise de circuitos elétricos e no uso de dispositivos como amperímetro, voltímetro, fonte elétrica e resistores. Algo similar foi feito por Santos & Dickman (10), que utilizaram experimentos reais com circuitos compostos por fonte e resistores e simuladores virtuais.

Abordagens diferenciadas também podem ser encontradas na literatura. Vasconcellos (2024) (11) aborda a Lei de Ohm a partir de analogias entre fenômenos hidráulicos e elétricos. Já Baião, Amaral & Veraszto (2017) (12) apresentam uma metodologia envolvendo o uso do Arduino combinado ao software Scratch. Castro (2019) (13) fez algo similar, realizando uma atividade totalmente virtual usando o Scratch. Domingos & Teixeira (2022) (14) usaram um simulador do PhET para a abordagem da lei. A utilização de simuladores com ênfase no Ensino à Distância também foi investigada por Silva & Duarte (2023) (15).

Estratégias e metodologias para tratar do tema de forma mais inclusivas também são encontradas na literatura. Como exemplo, podemos citar Velloso *et al*. (2021) (16), que desenvolveram atividades por meio da Aprendizagem Cooperativa no nível superior para deficientes visuais.

Diversos outros trabalhos utilizam-se da Lei de Ohm aplicada a medidas de grandezas físicas de camadas de grafite de lápis, tanto com intuito pedagógico como para realizar medidas experimentais. Propostas de atividade para a quantificação da espessura de grafite sobre um papel foram descritas para o Ensino Médio (17), sem levar em consideração tratamento estatístico adequado e coerente com o Ensino Superior. Rocha Filho et al. (18) propuseram uma atividade para análise e cálculo de resistência elétrica de associação de resistores em série e em paralelo construídos com diversos riscos fortes de lápis com grafite mole, tipo 6B, sobre papel sulfite para o Ensino Médio com ênfase na aprendizagem experimental de eletricidade, com auxílio de um multímetro. Vilela e Pereira (3) desenvolveram proposta similar, também voltada para alunos do Ensino Médio. Uma das conclusões dos autores foi a de que a técnica de perfilometria, utilizada para investigar a espessura dos traços, não se apresentou como uma solução eficaz e precisa para o problema, pois constatou-se que a rugosidade do próprio papel no qual o grafite foi depositado já se aproximava da ordem da espessura do traço de grafite.(19)

Estudos que analisam aspectos associados ao grafite depositado sobre papel ilustram outras aplicações e potencialidades relacionadas à proposta que será aqui relatada. Santiago et al. (2017) demonstraram que camadas de grafite provenientes de lápis do tipo 4B (General's Pencil Company, NJ, United States) depositadas manualmente sobre papel tratado com cera podem ser utilizadas na fabricação simples e rápida de sensores eletroquímicos de alto desempenho (20). Srinivas e Kumar reuniram trabalhos sobre a aplicação de eletrodos de grafites de lápis como sensores de dopamina (21). O artigo revisa o uso do eletrodo de grafite de lápis (EGL) como uma alternativa acessível e eficiente para aplicações eletroanalíticas, destacando sua aplicabilidade no ensino de conceitos eletroquímicos. Embora o EGL exija tratamentos de superfície para ativação devido à sua natureza eletro-inativa, ele se mostra eficaz para a detecção sensível de dopamina, um modelo comum em estudos. Desta forma, a pesquisa indica que o EGL, quando tratado corretamente, oferece uma ferramenta de baixo custo e alta seletividade para experimentos educacionais, possibilitando a introdução de técnicas de eletroquímica e análise de compostos em sala de aula. A revisão também sugere que futuras pesquisas sobre a funcionalização da superfície e a análise de dopamina em presença de outros analitos possam expandir ainda mais sua aplicabilidade em contextos educacionais e de pesquisa.

Dhanabalan, Chua e Ricardo (22) desenvolveram um estudo abrangente sobre as propriedades das camadas de diversos tipos de grafites (B até 8B) provenientes de lápis aplicados como eletrodos sobre papel, avaliando como algumas variáveis podem influenciar na resistência dos eletrodos de grafite depositados: a direção de deposição com relação à diferença de potencial; o comprimento e a área de seção das camadas depositadas; o número de camadas depositadas; o tempo de aplicação de diferença de potencial; e o tipo de grafite.

Grisales et al. (2016) (23) apresentaram um método simples e de baixo custo para a produção de tinta condutiva à base de grafite e sua aplicação na construção de circuitos elétricos em papel para aplicação pedagógica. A tinta demonstrou comportamento ôhmico e permitiu a criação de resistores e capacitores de placas paralelas com diferentes geometrias. Os resultados indicam que a atividade usando a tinta condutiva é uma ferramenta didática eficaz para o ensino de eletricidade e eletrônica, possibilitando a visualização experimental da relação entre resistência, capacitância e fatores geométricos dos circuitos.

Tendo em vista a literatura revisada, o presente trabalho tem como objetivo propor uma atividade experimental para disciplinas de Eletricidade no Ensino Superior. A proposta é investigar a aplicação da Lei de Ohm na aferição de espessura nanométrica de traços de grafite fazendo uso, entre outras coisas, da técnica de microscopia de varredura por sonda (*scanning probe microscopy* – SPM). Além disso, a facilidade de montagem do experimento, possibilita o maior engajamento dos alunos (2) e converge para uma abordagem mais investigativa (6).

As seções a seguir se concentram em realizar uma revisão teórica da Lei de Ohm; e sobre algumas propriedades do grafite, material que será utilizado no experimento.

### *Lei de Ohm*

A corrente elétrica que atravessa um material, devido a uma diferença de potencial entre dois pontos, pode se comportar de maneira diferente dependendo do tipo de material, mesmo sob as mesmas condições de tensão, área de contato elétrico e comprimento do percurso ao longo do material. Sabe-se que a corrente elétrica pode depender de diversos outros fatores tais como temperatura, pressão, estado físicos e composição do material (puro ou mistura). No entanto, apesar dos diversos fatores que podem influenciar na capacidade de condução dos materiais e diversas relações entre diferença de potencial aplicada (V) e a corrente elétrica (I), há um caso mais simples na qual há uma proporção direta essas duas últimas grandezas:

$$V \propto I \qquad (1)$$

Para esse caso, a constante de proporcionalidade, indicada por R, é denominada resistência elétrica. Ela é interpretada como a dificuldade que um meio imprime sobre o movimento de cargas elétricas (corrente elétrica) sob à ação de uma diferença de potencial. Logo:

$$V = R.I \tag{2}$$

A expressão matemática acima é creditada ao cientista alemão George Simon Ohm (1789 – 1854), que desenvolveu estudos na área, sendo conhecida como Lei de Ohm (24). A unidade de medida de resistência elétrica é definida como a razão entre a diferença de potencial elétrico e a corrente elétrica, isto é, volts por ampère (V / A) definida como Ohm (Ω) em homenagem ao cientista alemão (25).

Pode-se também definir uma grandeza vetorial que envolve a corrente elétrica em um região tridimensional, denominada como densidade volumétrica de corrente referenciada pela letra $\vec{J}$ (vetor "jota"). Existe um equivalente vetorial para a Equação ( 2 ) que envolve $\vec{J}$ definido por (26):

$$\vec{J} = \sigma.\vec{E} \tag{3}$$

Em que $\sigma$ é denominado como condutividade do meio ou condutância, com unidade de medida de Siemens por metro (S / m) e $\vec{E}$ é o campo elétrico aplicado, com unidade de medida de (N/ C ou V/m). A condutividade do meio é uma propriedade intrínseca ao material e diz respeito sobre a capacidade do material em conduzir corrente elétrica. A partir da condutividade do meio pode-se definir outra grandeza física que é o seu inverso, a resistividade ($\rho$) (26):

$$\rho = \frac{1}{\sigma} \tag{4}$$

Sendo sua unidade de medida definida como Ohm–metro (Ω.m) (26). Os valores de $\rho$ definem a classificação dos tipos de materiais quanto à capacidade de condução elétrica, sendo valores entre $5,0 \cdot 10^{-7}$ e $1,5 \cdot 10^{-8}$ Ω·m, considerados materiais condutores, valores típicos de metais e ligas metálicas; valores entre $5,0 \cdot 10^{-6}$ a $3,3 \cdot 10^{6}$ Ω·m, considerados materiais semicondutores; e valores acima da ordem de $10^{-10}$ Ω·m, considerados isolantes. (27)

A relação entre corrente elétrica e densidade de corrente pode ser estabelecida pela seguinte relação (28):

$$I = \int \vec{J} \cdot \vec{dA} \tag{5}$$

Em que $\vec{dA}$ é um elemento infinitesimal de superfície orientada, cuja a direção e sentido são definidos pelo vetor normal à superfície com módulo igual ao valor da área. A Figura 1a esquematiza a Equação ( 5 ).

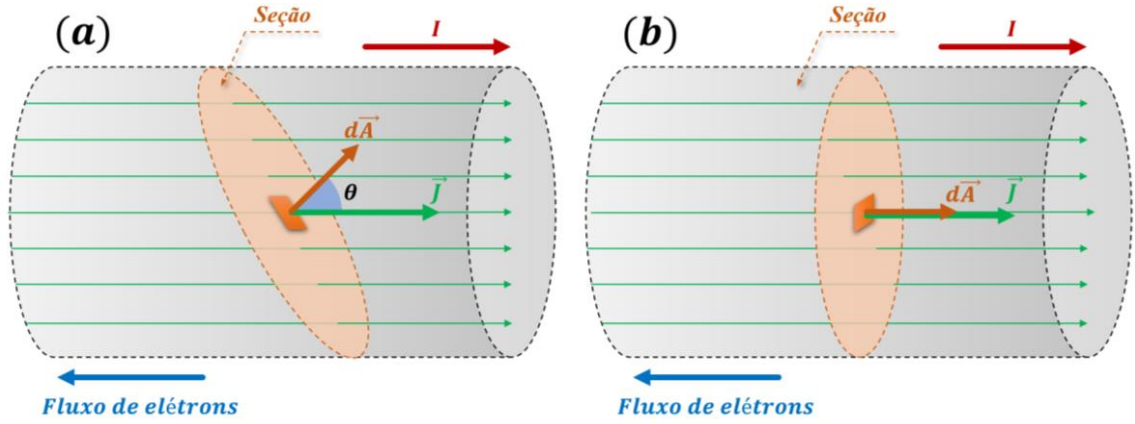

**Figura 1.** Representação da relação entre corrente elétrica, densidade de corrente e superfície orientada para duas situações distintas: (a) quando $\vec{J}$ e $\vec{dA}$ não são paralelos; (b) quando $\vec{J}$ e $\vec{dA}$ são paralelos.

Para uma situação específicada na qual J é paralelo aos elementos infinitesimais $\vec{dA}$ (Figura 1 b), pode-se considerar que (28):

$$I = \int \vec{J} \cdot \vec{dA} = \int J.dA.\cos(0°) = J\int dA = J \cdot A \qquad (6)$$

Em que a integral do elemento infinitesimal de área $dA$ tem como valor a área da superfície de análise do fluxo.

Para obter outra relação importante a respeito da resistência e resistividade de um material, considerar-se-á um cilindro com corte transversal formando uma seção de corte de área A e com comprimento L feito de um material com condutividade $\sigma$. Considerando que o potencial é constante e homogêneo em suas extremidades de maneira a formar um campo elétrico perpendicular à área da seção, a partir do módulo da Equação ( 3 ) pode-se fazer as seguintes relações com a Equação ( 3 ) e ( 6 ):

$$I = J \cdot A = \sigma.E.A \qquad (7)$$

Sabendo que o campo elétrico pode ser definindo em termos da diferença de potencial (V) e o caminho (L) no qual o campo elétrico se estende paralelamente (29):

$$E = \frac{V}{L} \qquad (8)$$

Logo, substituindo as Equações ( 4 ) e ( 8 ) em ( 7 ) e escrevendo V em função de I (26,29):

$$V = \frac{\rho.L}{A}I \qquad (9)$$

Comparando a Equação ( 9 ) com ( 1 ), pode-se encontrar um relação da resistência de um material em função de suas propriedades geométricas como comprimento (L) e área da seção do material (A) e propriedades composicionais como a resistividade ($\rho$), sendo representada esquematicamente pela Figura 2. Logo (29):

$$R = \frac{\rho.L}{A} \qquad (10)$$

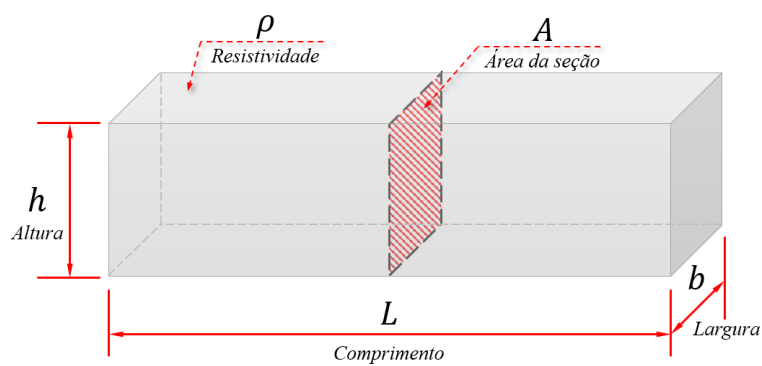

**Figura 2.** Representação esquemática da Equação ( *10* ).

Para deduções e abordagens mais detalhadas, recomenda-se a leitura dos livros de Kleber Daum Machado (25) e David J. Griffiths (26).

Vale notar que em livros e em aulas de Física do Ensino Médio a relação (11) costuma ser chamada de 2ª Lei de Ohm, enquanto a Equação ( 2 costuma ser nomeada como a 1ª Lei de Ohm.

Os materiais que apresentam uma relação linear entre tensão elétrica aplicada e corrente elétrica são denominados como mateirais ôhmicos, enquanto que um material não ôhmico apresentará uma relação não linear entre essas duas grandezas. (27)

### *As propriedades do grafite*

O grafite é uma forma estável do carbono em condições ambientais de temperatura e pressão, sendo sua estrutura cristalina composta por planos paralelos formados por átomos de carbono dispostos nos vértices de hexágonos regulares interligados. No interior dos planos formados, denominados como camadas ou lâminas, cada átomo de carbono se conecta a três vizinhos coplanares por meio de ligações envolvendo orbitais híbridos $sp^2$. Vale ressaltar que cada uma dessas camadas individuais corresponde ao grafeno, um nanomaterial de carbono com propriedades excepcionais. O primeiro grafeno obtido foi produzido a partir da separação sucessiva de camadas de uma peça de grafite, utilizando uma fita adesiva plástica, até restar apenas uma única camada de átomos de carbono. Esse método, conhecido como esfoliação micromecânica ou técnica da fita adesiva, foi um dos primeiros a permitir a obtenção de grafeno de alta qualidade. A configuração hexagonal característica das ligações $sp^2$ pode ser observada na Figura 3. Além disso, o quarto elétron de valência de cada átomo não está restrito a uma ligação específica, mas sim deslocalizado, contribuindo para um orbital molecular que se estende ao longo da estrutura e se situa entre as camadas. As interações entre essas camadas ocorrem perpendicularmente aos planos, ao longo da direção cristalográfica c, conforme indicado na Figura 3, e são do tipo van der Waals, forças relativamente fracas que mantêm as lâminas unidas. (27)

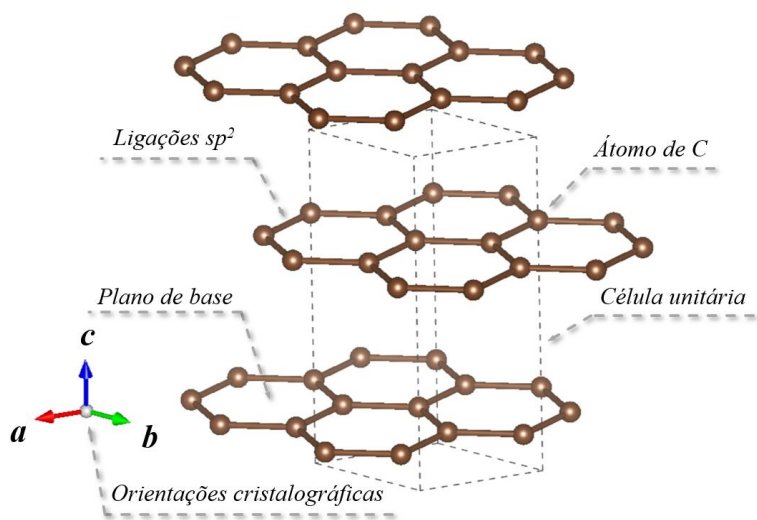

**Figura 3.** Estrutura cristalina do grafite. Desenhado com auxílio do software gratuito *VESTA (30)*.

O grafite apresenta características tanto metálicas quanto não metálicas, o que o torna um material valioso para a fabricação de eletrodos (31). Em sua forma pura, o grafite apresenta resistividade entre $5,0 \cdot 10^{-6}$ e $3,3 \cdot 10^{-5}$ $\Omega \cdot m$ (27). O lápis de grafite é produzido a partir de uma combinação de materiais, onde o pó fino de grafite é incorporado a uma matriz que pode ser inorgânica, como resinas, ou orgânica, incluindo argila ou polímeros de alto peso molecular, como a celulose (32). Basicamente, os lápis de grafite são compostos por uma mistura de aproximadamente 65% de grafite, 30% de argila e uma fração de agentes aglutinantes, como ceras, resinas ou polímeros (33). Conforme escala europeia estabelecida, os lápis de grafite são categorizados por letras e números em função de sua composição, sendo que a letra H indica dureza (hardness), apresentando maior quantidade de argila, conferindo-lhes maior dureza; e a letra B representa a tonalidade escura (blackness), possuindo maior quantidade de grafite, tornando-se mais macios. Conforme a escala, os lápis possuem graduações que variam de 9H (o mais duro) até 9B (o mais macio), sendo o lápis do tipo HB composto por uma proporção equilibrada de grafite e argila (34,35,36,37,38). A presença de argila na composição dos lápis influencia diretamente suas propriedades químicas, como a capacidade de troca iônica, e estruturais, incluindo o grau de desordem e a morfologia da superfície (36).

Devido à sua estrutura cristalina, o grafite exibe propriedades fortemente associadas à anisotropia, ou seja, suas propriedades variam dependendo da direção cristalográfica ao longo da qual são medidas. Com relação às propriedades elétricas, as resistividades elétricas nas direções paralela e perpendicular ao plano do grafeno, as lâminas ou camadas, são, respectivamente, da ordem de $10^{-5}$ e $10^{-2}$ $\Omega.m$. Esse fenômeno pode ser explicado pela alta mobilidade dos elétrons livres no grafite, cujos movimentos são mais favorecidos ao serem submetidos à ação de um campo elétrico aplicado na direção paralela ao plano, o que resulta na sua resistividade relativamente baixa nessa direção (27). Além do mais, devido às fracas ligações de van der Waals entre as camadas, os planos podem deslizar facilmente uns sobre os outros, o que confere ao grafite suas excelentes propriedades de lubrificação (27) e explica o uso como lápis de escrever. Mecanicamente, o grafite é muito macio e quebradiço, possuindo um módulo de elasticidade relativamente menor com relação ao diamante, outro polimorfo do carbono.

Sua condutividade elétrica no plano é de $10^{16}$ a $10^{19}$ vezes maior que a do diamante, enquanto a condutividade térmica é praticamente a mesma. O coeficiente de expansão térmica do grafite, no plano, é pequeno e negativo, enquanto na direção perpendicular ao plano é positivo e relativamente grande, ao contrário do diamante, que possui um coeficiente de expansão térmica pequeno e positivo. Opticamente, o grafite é opaco e apresenta uma coloração negro-prateada (27).

Outras propriedades importantes do grafite incluem boa estabilidade química em altas temperaturas e em atmosferas não oxidantes, alta resistência ao choque térmico, elevada adsorção de gases e boa usinabilidade. Devido às suas diversas propriedades, as suas aplicações são bem diversificadas, como: lubrificantes, eletrodos de baterias, materiais de fricção (como sapatas de freio), resistências para fornos elétricos, eletrodos de solda, cadinhos metalúrgicos, refratários e isolantes para altas temperaturas, tubeiras de foguetes, vasos de reatores químicos, contatos elétricos e dispositivos de purificação do ar (27).

# METODOLOGIA

A metodologia utilizada foi dividia em cinco subseções: determinação da resistividade do grafite 6B; deposição de camada de grafite sobre papel vegetal; medição da largura média da camada de grafite; determinação da espessura da camada de grafite depositada; e medição da espessura média da camada de grafite usando o SPM.

## *Materiais*

Para a realização das medidas de resistividade do grafite foram utilizados um lápis grafite ecolápis castell 9000 6B sextavado Faber-Castel, papel vegetal Canson A4 60 g/m$^2$, papel sulfite A4 75 g/m$^2$ Chamex branco, estilete, pedra de granito polido para apoio, régua de 30 cm milimetrada, caneta esferográfica, dois multímetros de bancada POL-79C Dual Display Multimeter, fonte de alimentação Instrutherm modelo FA-3050 simétrica digital de 2 canais, 2 pinos tipo "banana", 2 garras tipo "jacaré", paquímetro digital Marberg 0 - 200 mm, 2 pinças (garra) de 3 dedos com mufa fixa, 2 mufas duplas simples, suporte universal, prancheta de plástico e microscópio de varredura por sonda Shimadzu SPM-9700HT.

## *Métodos*

### *Determinação da resistividade do grafite 6B*

O sistema de medição foi construído conforme a Figura 4.

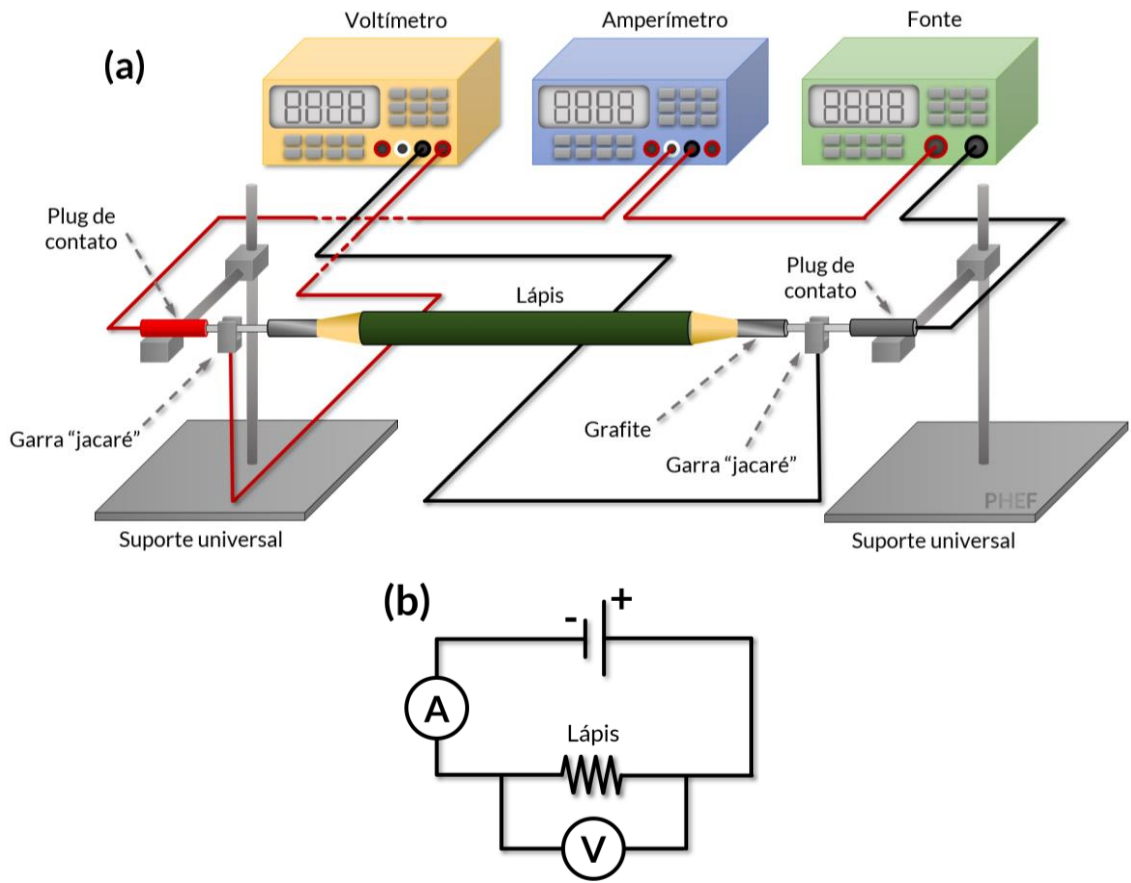

**Figura 4.** (a) Esquema da montagem para realizar a medição da resistividade. (b) Representação do circuito elétrico utilizado para realizar a medição.

    Inicialmente as extremidades do lápis foram descascadas com o auxílio do estilete. Com o auxílio do paquímetro digital, foram realizadas dez medidas do diâmetro de cada lado do grafite, totalizando 20 medições ao todo. Essas medidas foram usadas para calcular o diâmetro médio do grafite e, consequentemente, a área de seção transversal do grafite por onde a corrente elétrica passou. Em ambos os lados, a seção do lápis foi desbastada utilizando um estilete e riscando-o sobre uma folha de papel sulfite A4 (75 g/m², Chamex branco) apoiada em uma superfície lisa, de modo a obter uma face plana o mais perpendicular possível ao eixo longitudinal do lápis. Na sequência, utilizou-se o paquímetro para realizar a medida da dimensão longitudinal do grafite do lápis, medida esta considerada o comprimento (L). Nas extremidades do grafite foram posicionados plugs de contato tipo "banana" que desempenharam o papel de terminais elétricos para a fonte e o multímetro utilizado na função amperímetro. Fez-se necessário realizar uma pressão significativa entre o plug e o grafite e verificar se havia passagem de corrente. Os terminais do multímetro com função de voltímetro foram conectados às extremidades do grafite com o auxílio de plugs do tipo "jacaré", sendo conectados aos terminais do voltímetro. Com o sistema montado conforme a Figura 5, foram medidas as correntes elétricas que passavam pelo grafite conforme a tensão entre os terminais era aumentada.

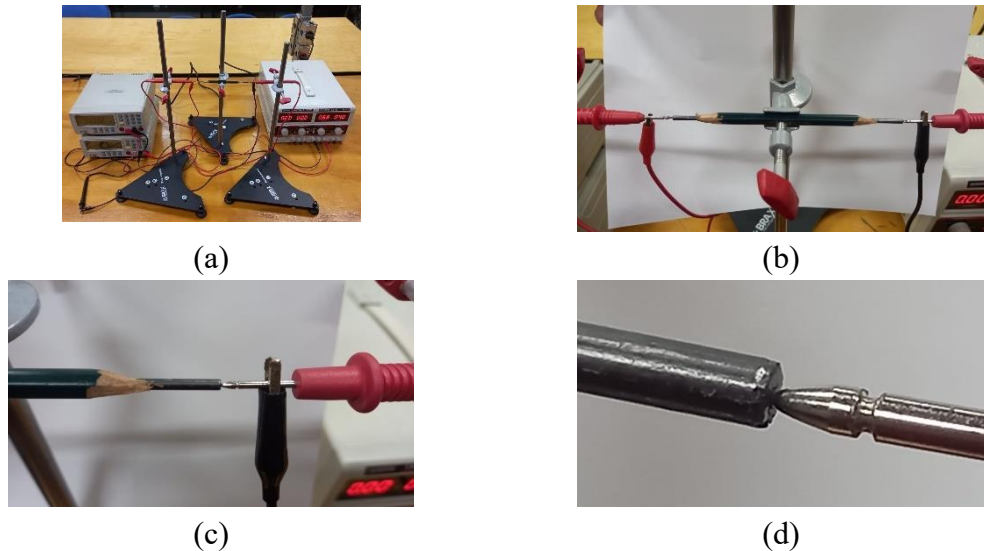

(a)                  (b)

(c)                  (d)

**Figura 5.** (a) Fotografia da montagem experimental utilizada. (b) Fotografia detalhando o contato dos plugs com o grafite do lápis. (c) Fotografia detalhando contato dos plugs do amperímetro e do voltímetro com o grafite. (d) Fotografia detalhando o contato do plug tipo "banana" com a seção perpendicular ao eixo longitudinal do grafite.

Para o estudo das grandezas físicas envolvidas no experimento, foi mais coerente escrever a Equação ( 9 ) com I em função de V, já que a corrente elétrica aplicada no material varia conforme a diferença de potencial aplicada entre os terminais do lápis é variada:

$$I = \frac{A}{\rho.L}.V \qquad (11)$$

Desta forma, pode-se utilizar o método dos quadrados mínimos para obter uma função linear que descreve a relação entre as grandezas envolvidas na Equação ( 12 ):

$$f(x) = a.x + b \qquad (12)$$

Em que **a** é o coeficiente angular, **b** é o coeficiente linear e **x** a variável independente. Comparando com a Equação ( 11 ) nota-se que **a** está associada ao valor de V aplicado, as grandezas físicas intrínsecas ao material (área da seção, comprimento e resistividade) desempenhando o papel de coeficiente angular da função linear (a) e f(x) está associado ao valor de I, o qual depende de V. O valor do coeficiente linear deverá se aproximar de 0, pois trata-se de grandezas diretamente proporcionais. Desta forma, a resistividade do grafite 6B pode ser obtida a partir da seguinte relação:

$$a = \frac{A}{\rho.L} \qquad (13)$$

Isolando ρ:

$$\rho = \frac{A}{a.L} \qquad (14)$$

O valor de ρ obtido será usado na etapa posterior, descrita na próxima seção.

### *Deposição da camada de grafite sobre papel vegetal*

A deposição da camada de grafite foi realizada com o mesmo grafite da seção anterior sobre papel vegetal Canson A4 60 g/m² translúcido apoiado sobre pedra de granito polido. O modo de deposição foi realizado com o lápis em posição perpendicular com relação ao papel vegetal aplicando-se uma pressão com a mão suficiente para realizar um risco contínuo e homogêneo, conforme a Figura 6. Foram depositadas 50 camadas com o movimento em uma única direção do lápis, isto é, ao chegar ao final do risco, o lápis foi levantado e posicionado novamente na posição inicial para iniciar a realização do risco na mesma direção. Recomenda-se realizar um risco de 12 cm de comprimento, 2 cm a mais do que os 10 cm que serão utilizados para as medições. A fim de auxiliar nas medidas posteriores, foram realizadas marcações de distâncias de 0 a 10 cm, com espaçamento de 2 em 2 cm, com auxílio de uma régua de 30 cm milimetrada (Erro: ± 0,05 cm) e caneta esferográfica. Vale ressaltar que um teste empírico e rápido para avaliar a efetividade da deposição consiste em direcionar a folha a uma fonte de luz e observar se há dificuldade na passagem da luz através do grafite.

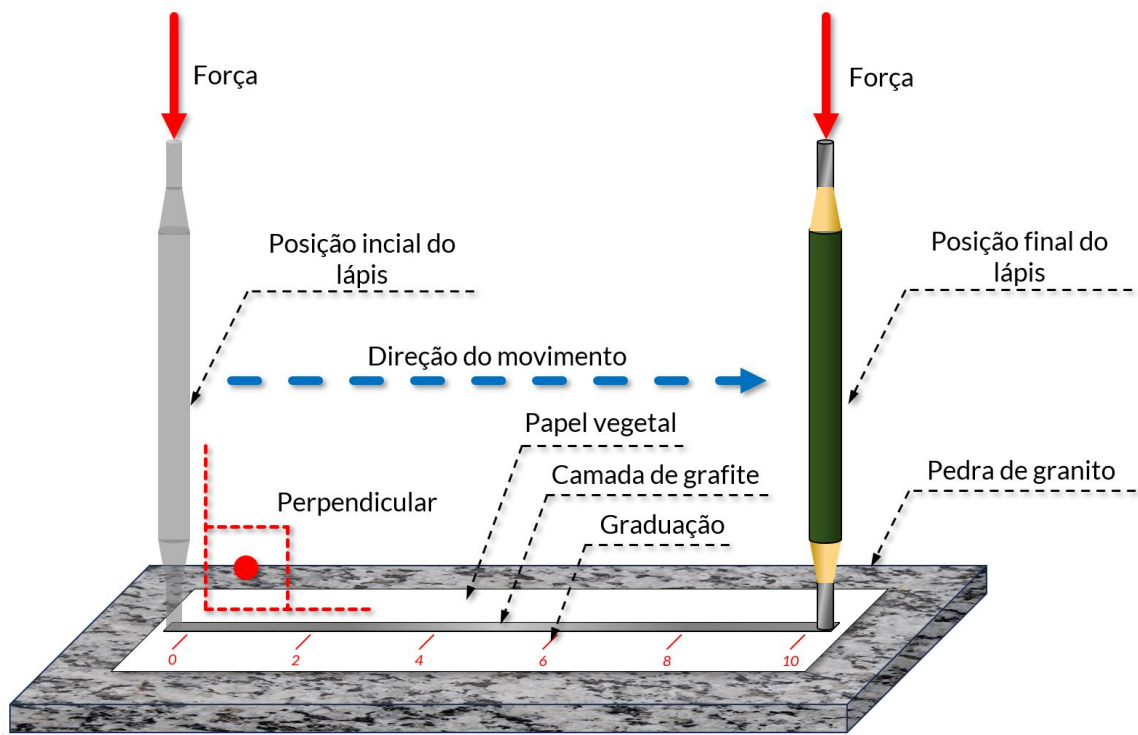

**Figura 6.** Método de deposição da camada de grafite.

### *Medição da largura média da camada de grafite*

As medidas da largura da camada de grafite depositada sobre o papel vegetal, também denominada como a largura (b) na Figura 2, foram realizadas com auxílio da câmera traseira do celular Samsung A12 de resolução 8000 x 6000 pixels. O celular foi posicionado sobre a camada de grafite depositada no papel a partir de um suporte universal com o auxílio de mufas. As larguras do traço de grafite ao longo de sua extensão foram medidas utilizando-se o software *ImageJ*, tomando-se como referência a escala do paquímetro Marberg aberto em 5,00 mm, posicionado próximo à camada de grafite. Realizou-se 50 medidas de cada um dos diferentes trechos: 0 a 2 cm; 2 a 4 cm; 4 a 6 cm; 6 a 8 cm; e 8 a 10 cm. A Figura 7 b e c demonstram exemplos de imagens coletadas.

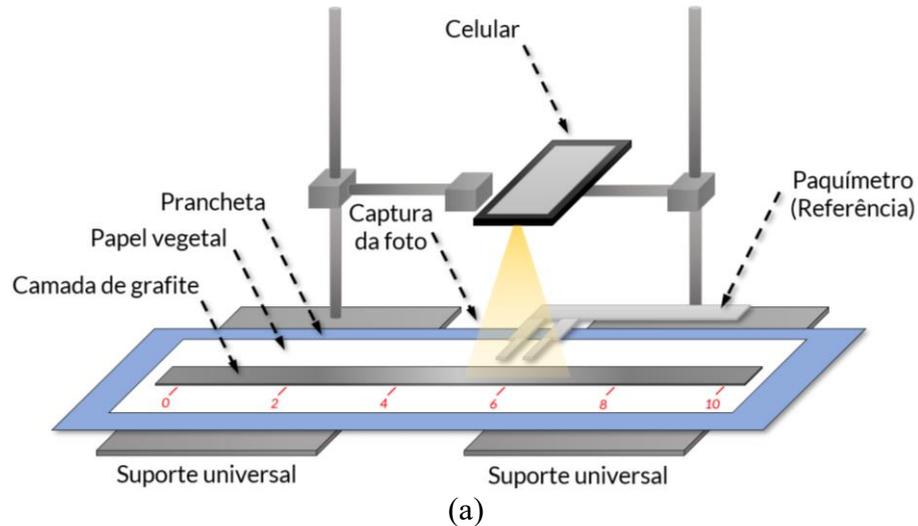

(a)

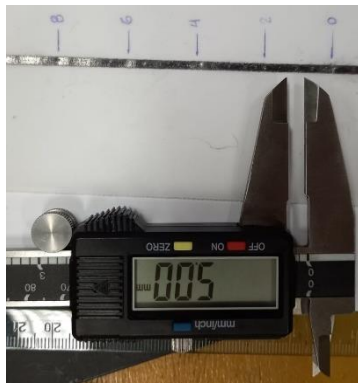 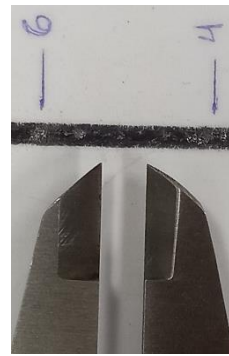

(b) (c)

**Figura 7.** (a) Esquema da montagem experimental para obter imagens da camada de grafite. (b) Paquímetro Marberg posicionado próximo à camada de grafite depositada. (c) Zoom na região de 4 a 6 cm da camada de grafite, conforme graduação previamente realizada.

### *Determinação da espessura da camada de grafite depositada*

A espessura do grafite pode ser determinada utilizando-se as relações desenvolvidas na seção "Lei de Ohm". A área da seção da camada de grafite pode ser aproximada para um retângulo de largura b e altura h, em que h é o valor que se deseja obter, no caso a espessura da camada de grafite, conforme a Figura 2. O produto de b por h resulta na área da seção (A), desta forma, usando a Equação ( 3 ):

$$I = \frac{b.h}{\rho.L}.V \qquad (15)$$

Realizou-se medidas da corrente elétrica (I) que passa pela camada de grafite em função da diferença de potencial (V) aplicada em seus terminais para diferentes valores de L: 2, 4, 6, 8 e 10 cm, conforme graduação auxiliar feita no papel. As distâncias foram ajustadas com relação à distância entre os dois pinos tipo banana em contato com a camada de grafite centralizada, isto é, posicionando-os longe das extremidades transversais, a fim de manter maior homogeneidade na medida, já que as extremidades podem apresentar baixa deposição de camadas de grafite. A Figura 8 mostra o esquema de motagem do sistema de medição e a Figura 9 (a) mostra o esquema real. A Figura 9 (b) mostra o esquema do circuito, no qual a camada de grafite atua como um potenciômetro devido à variação do valor de sua resistência conforme a variação do comprimento medido. Com o objetivo de aprimorar o contato entre a camada de grafite e os plugs, inseriu-se uma folha de papel sulfite A4 (75 g/m², Chamex branco) sob o papel vegetal. As Figura 9 (c) e (d) mostram o esquema real. A Figura 9 (d) mostra em detalhe como o contato entre o plug e a camada de grafite foi realizado.

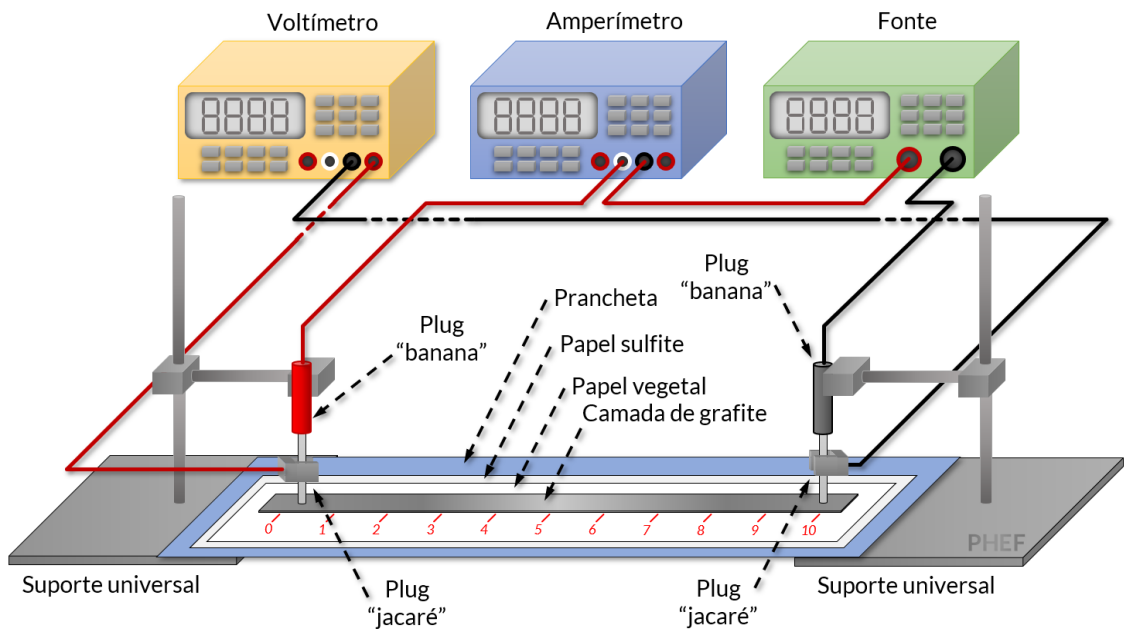

**Figura 8.** Esquema para medidas elétricas na camada de grafite depositada sobre o papel vegetal.

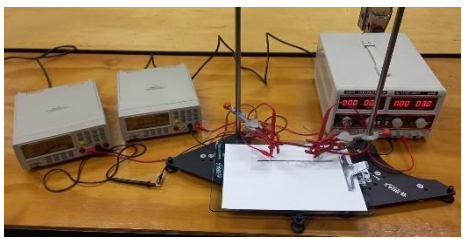

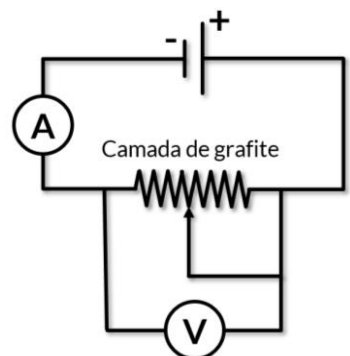

(a)                    (b)

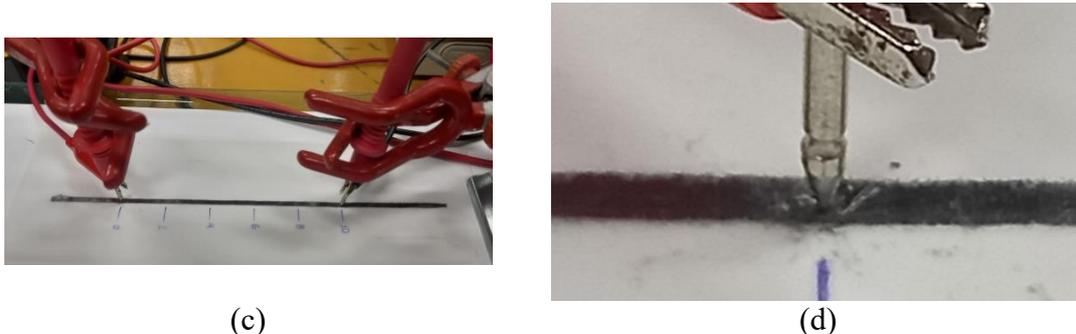

|       (c)       |       (d)       |

**Figura 9.** (a) Circuito elétrico representando o sistema de medida; (b) Esquema para medidas elétricas na camada de grafite depositada sobre o papel vegetal. (c) Detalhe da conexão dos plugues tipo "banana" com a camada de grafite depositada sobre papel vegetal. (d) Detalhe do contato do plug tipo "banana" com a camada de grafite depositada sobre papel vegetal.

O valor de ρ a ser utilizado é o obtido por meio da consecução da primeira etapa experimental. Desta forma, pode-se dizer que o valor de h pode ser obtido comparando a Equação ( 15 a uma função linear (Equação ( 12) e utilizando novamente o método dos quadrados mínimos para obter os coeficientes da função linear, os quais são:

$$a = \frac{b.h}{\rho.L} \quad (16)$$

Com o valor de L sendo utilizado adequadamente para cada tomada experimental realizada em determinado comprimento L. Assim, com os cinco conjuntos de medidas obtidos conforme a metodologia descrita na seção "Medição da largura da camada de grafite", pode-se obter o valor de h, isolando essa grandeza:

$$h = \frac{\rho.L.a}{b} \quad (17)$$

Como a camada de grafite não é perfeitamente lisa nem possui seção transversal retangular, o valor obtido representa apenas uma aproximação, considerando a camada como um formato de paralelepípedo. Dessa forma, a partir dos diferentes valores de h determinados para distintos comprimentos *L*, pode-se calcular a média aritmética de h, visando estimar um valor representativo da espessura média da camada de grafite ao longo de toda a sua extensão.

### *Medição da espessura média da camada de grafite usando o SPM*

As medidas de espessura das camadas de grafite depositadas foram realizadas utilizando um microscópio de varredura por sonda (SPM) Shimadzu SPM-9700HT disponível numa instituição parceira, cujos erros instrumetais são de 0,2 nm no plano "xy" de varredura e 0,01 nm na altura (eixo z). A análise foi realizada utilizando uma ponta do tipo sem contato para medir a distribuição topográfica da interface entre papel vegetal e grafite. As imagens foram tratadas no *software* livre Gwyddion, versão 2.68. Os valores obtidos pelo SPM foram comparados com os valores obtidos por intermédio de cálculos feitos com base na Lei de Ohm e nos valores obtidos nas etapas anteriores.

# RESULTADOS

Os dados, as equações e os cálculos de propagação de erros utilizados no tratamento estatístico estão nos Apêndices.

### *Determinação da resistividade do grafite 6B*

Conforme a metodologia apresentada, relizou-se a medida do diâmetro médio do grafite do lápis obtendo-se o valor de 2,808(11) · $10^{-3}$ m e, consequentemente, uma área da seção do lápis de 6,19(5) · $10^{-6}$ m² (vide Apêndice A). O comprimento do grafite do lápis foi medido como sendo de 1,3915(1) · $10^{-1}$ m. A Figura 10 mostra o gráfico com os dados da corrente em função da tensão aplicada no grafite. A função obtida no ajuste dos quadrados mínimos para a relação foi:

$$I(V) = 2.395(16) \cdot 10^{-1} \cdot V - 1{,}42(32) \cdot 10^{-2} \qquad (18)$$

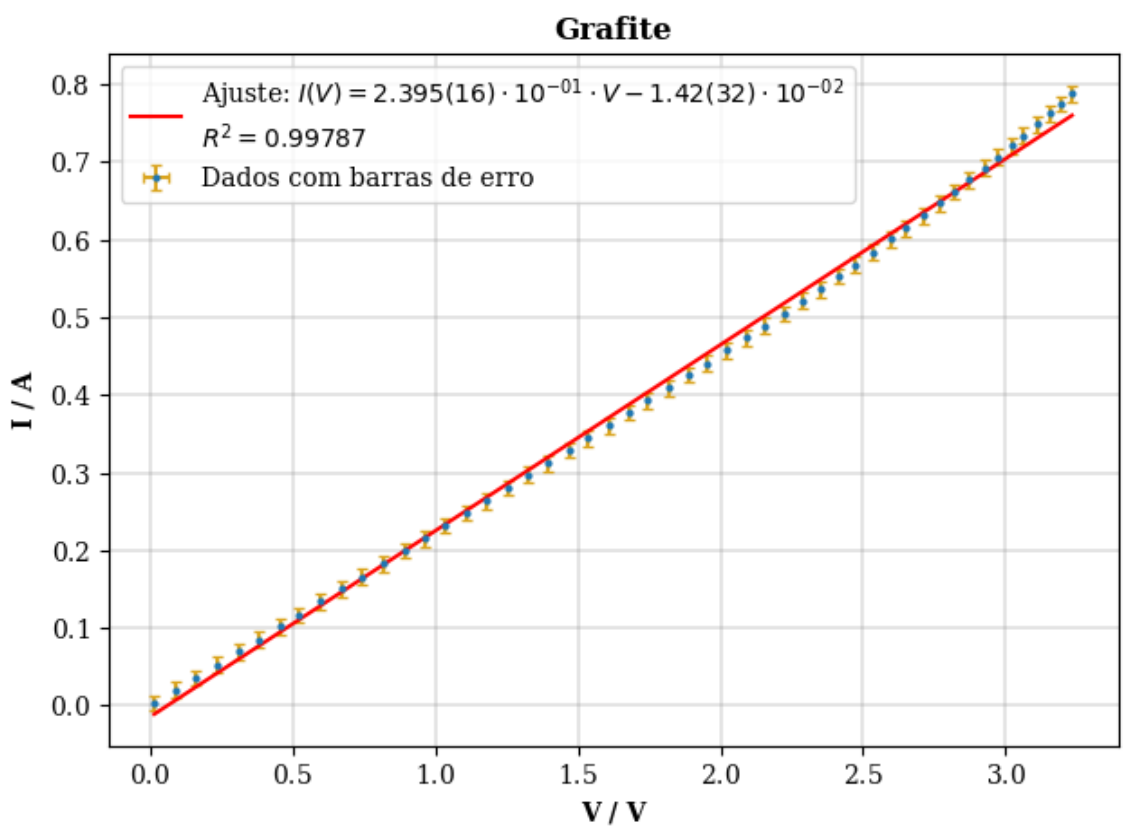

**Figura 10.** Gráfico de I em função de V para o grafite.

Em que o coeficiente angular **a** foi de 2,395(16) · $10^{-1}$ A/V, o coeficiente linear **b** foi de 1,42(32) · $10^{-2}$ A e o coeficiente de determinação ($R^2$) foi de 0,99787. Com os valores obtidos e realizando a propagação de erro adequada foi possível calcular o valor da resistividade do grafite do lápis a partir da Equação ( 14, obtendo-se o valor de 1,857(19) · $10^{-4}$ Ω·m. que se difere do valor esperado de 5,0 · $10^{-6}$ a 3,3 · $10^{-5}$ Ω·m do grafite puro devido à presença de outras substâncias em sua composição (vide Apêndice B). O valor obtido nesta etapa foi utilizado para o cálculo da altura da camada de grafite.

### *Medição da largura média da camada de grafite*

Após realizar a deposição das camadas de grafite sobre papel vegetal utilizando a metodologia descrita na seção "Deposição da camada de grafite sobre papel vegetal", realizou-se as medidas da largura (b) das camadas de grafite depositadas obtendo-se valores diferentes para os trechos escolhidos, demonstrados na tabela a seguir:

**Tabela 1.** Valores obtidos para as medidas da largura da camada de grafite depositada sobre o papel vegetal.

| Comprimento de Referência (cm) | Largura média (m) | Número de dados (N) |
|---|---|---|
| 0 - 2 | $2,390(23) \cdot 10^{-3}$ | 50 |
| 0 - 4 | $2,291(24) \cdot 10^{-3}$ | 100 |
| 0 - 6 | $2,342(20) \cdot 10^{-3}$ | 150 |
| 0 - 8 | $2,364(18) \cdot 10^{-3}$ | 200 |
| 0 - 10 | $2,351(18) \cdot 10^{-3}$ | 250 |

Realizou-se 50 medidas de diferentes trechos do grafite depositado. Isto é, para o trecho "0 – 2 cm" obteve-se 50 medidas da largura, para o trecho "2 – 4 cm" obteve-se outras 50 medidas e assim sucessivamente até o trecho "8 – 10 cm". Desta forma, os dados foram sendo agregados para fazer o tratamento estatístico, conforme o comprimento foi mudando (vide Apêndice C). Exemplo, o comprimento "0 – 6" envolve todos os dados dos trechos "0 – 2 cm", "2 – 4 cm" e "4 – 6 cm", totalizado 150 dados, já que se trata de 3 trechos. Os valores apresentados na Tabela 1 foram utilizados nos cálculos da seção a seguir.

### *Determinação da espessura da camada de grafite depositada*

Com os valores de resistividade ($\rho$) e largura da camada (b) de grafite depositada sobre a folha vegetal, obtidas a partir das metodologias descritas anteriormente, foi possível estimar as alturas de cada camada depositada de acordo com os diferentes comprimentos utilizando o valor do coeficiente angular obtido do ajuste, conforme Equação ( 17 . Os valores são apresentados na Tabela 2. Os dados e ajustes de retas são apresentados na Figura 11 e os detalhes dos cálculos apresentados no Apêndice D.

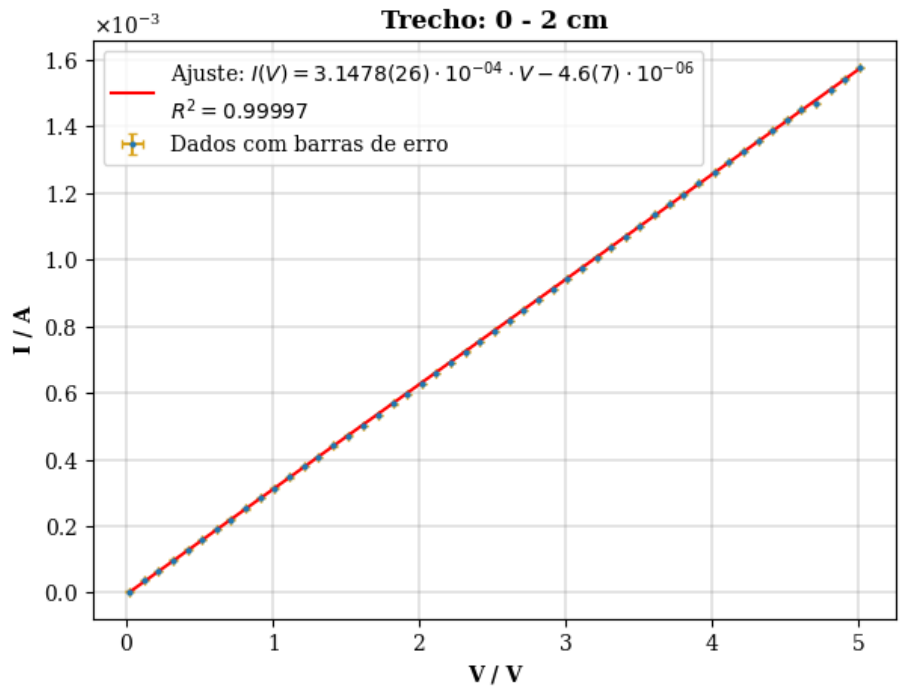

(a)

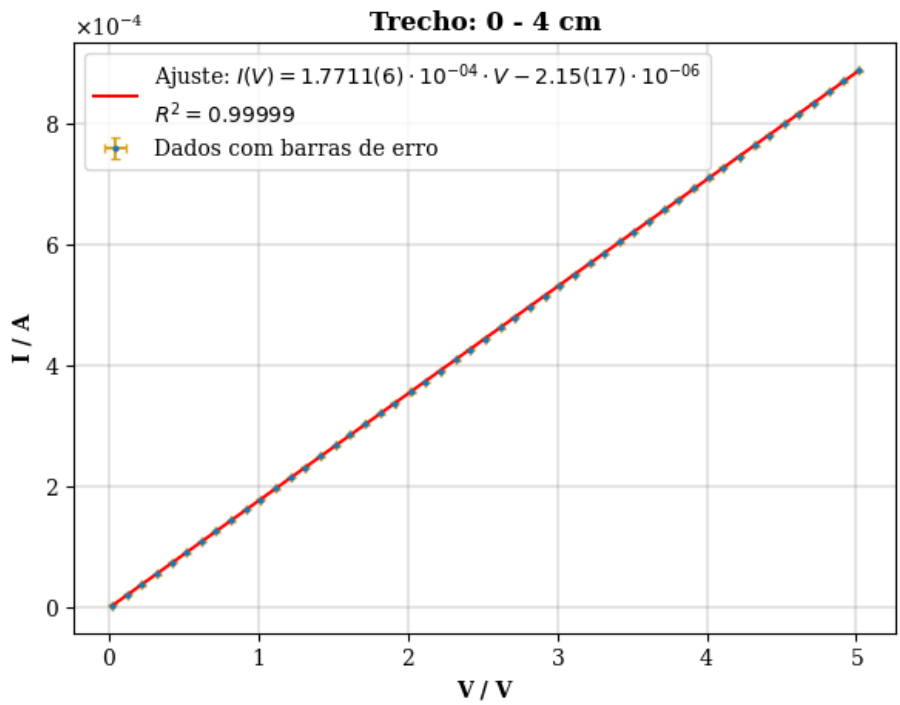

(b)

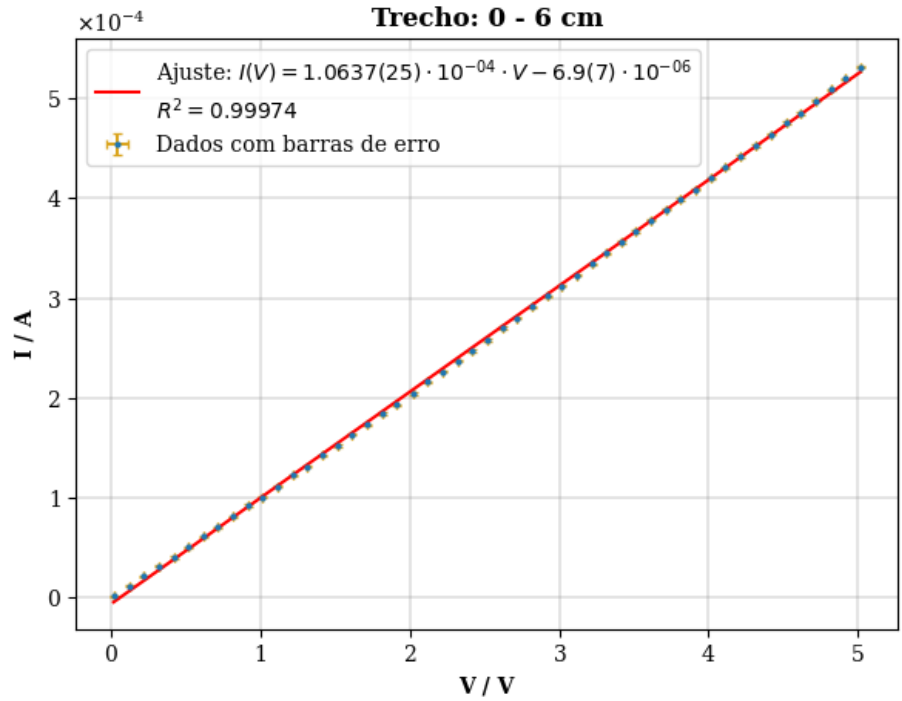

(c)

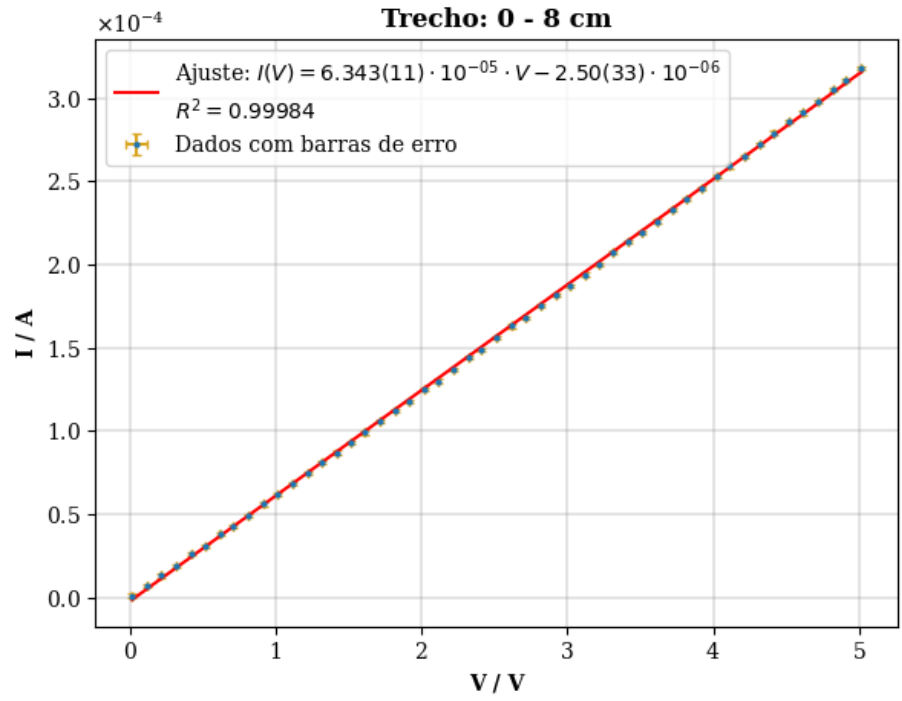

(d)

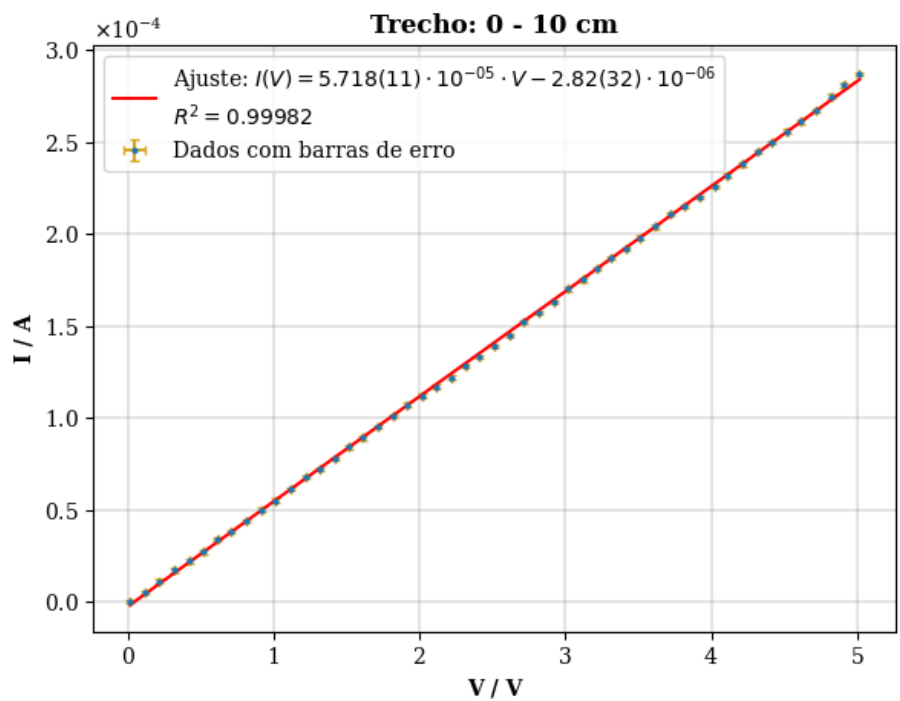

(e)

**Figura 11.** Gráficos obtidos a partir de cada comprimento de referência: (a) 0 – 2 cm; (b) 0 – 4 cm; (c) 0 – 6 cm; (d) 0 – 8 cm; (e) 0 – 10 cm.

**Tabela 2.** Medidas da altura da camada de grafite depositida sobre o papel vegetal para os diferentes comprimentos de referência.

| Comprimento de Referência (cm) | Altura da camada, h (nm) |
|---|---|
| 0 - 2 | 489(14) |
| 0 - 4 | 574(11) |
| 0 - 6 | 506(8) |
| 0 - 8 | 399(6) |
| 0 - 10 | 452(6) |
| **Média** | 456,5(34) |

*Medição da espessura média da camada de grafite usando o SPM*

A partir das análises pelo SPM foi possível obter imagens que demonstraram a presença de degraus em regiões de interface entre a camada de grafite depositada e o papel vegetal. As imagens são demonstradas nas Figuras Figura *12*, Figura **13** e Figura **14**.

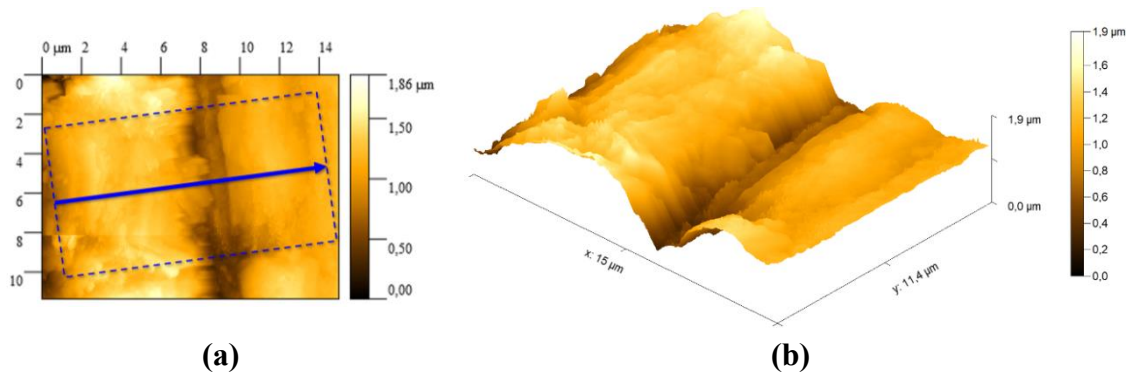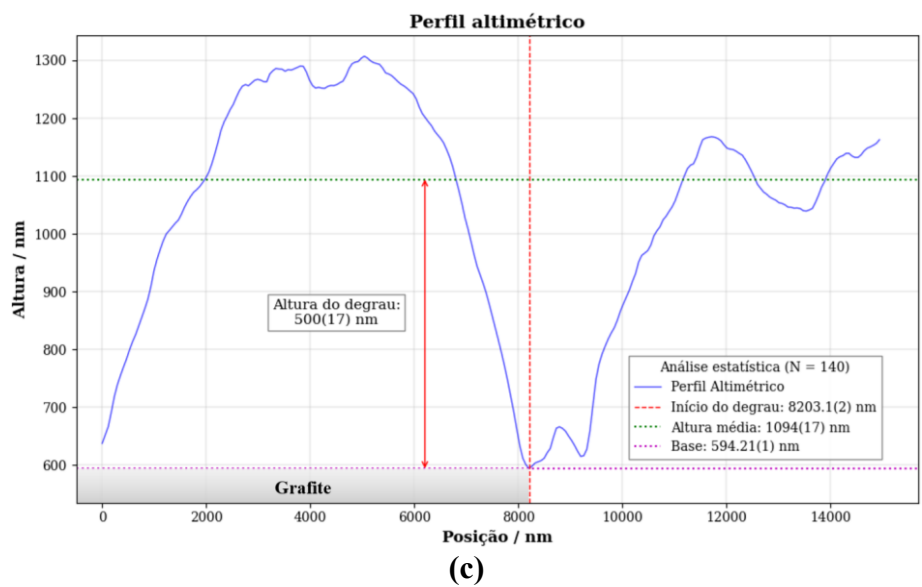

**Figura 12.** Figura de SPM obtida a apartir das análises de SPM: (a) Imagem bidimensional, o retângulo tracejado em azul indicaa área da qual foi extraído o perfil altimétrico médio e a seta indica o sentido correspondente no gráfico, sendo o início da seta a lateral esquerda do gráfico e a seta a lateral direita; (b) Imagem trimensional; (c) Perfil altimétrico médio da região destacada na imagem bidimensional.

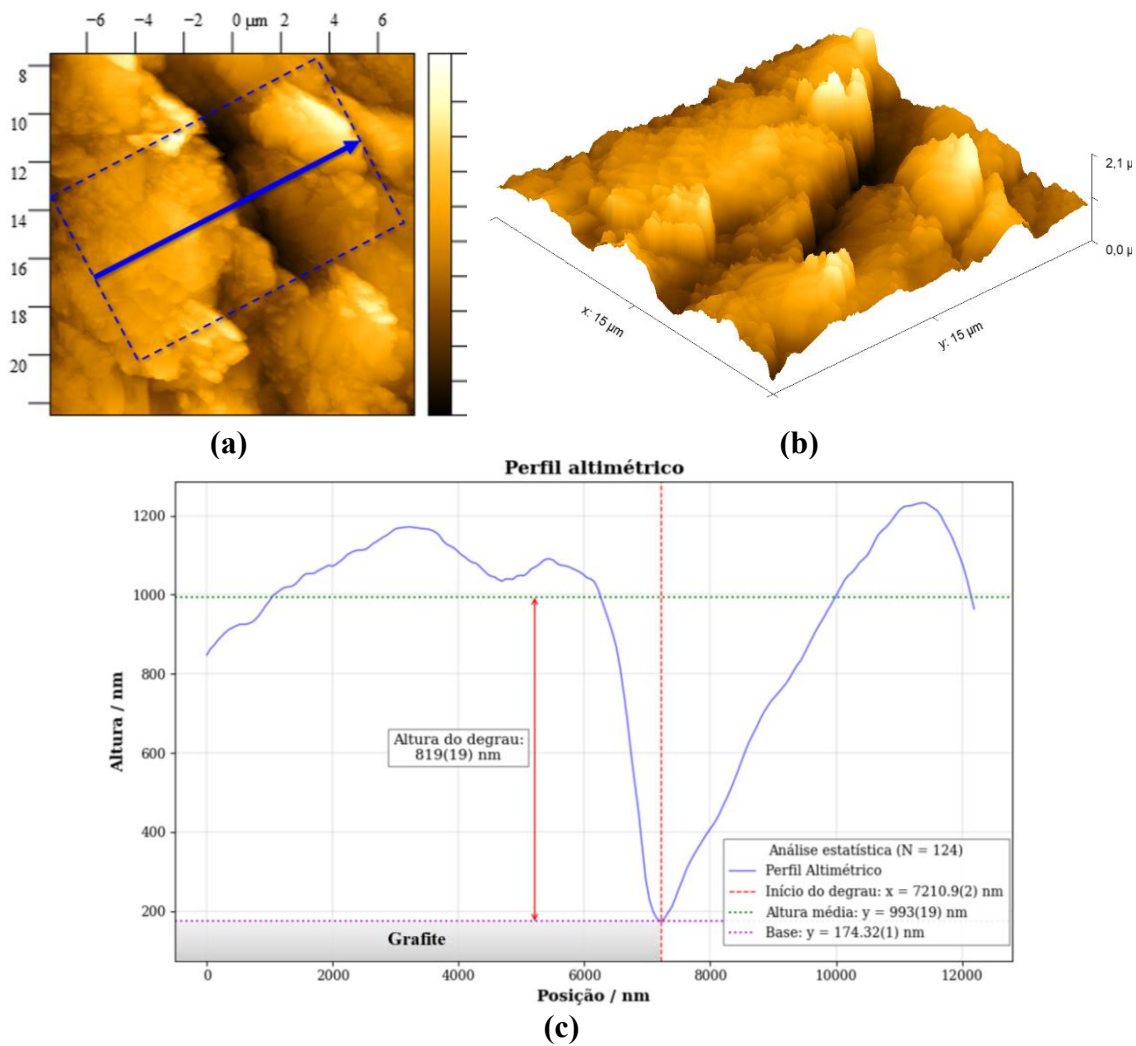

**Figura 13.** Figura de SPM obtida a apartir das análises de SPM: (a) Imagem bidimensional, o retângulo tracejado em azul indicaa área da qual foi extraído o perfil altimétrico médio e a seta indica o sentido correspondente no gráfico, sendo o início da seta a lateral esquerda do gráfico e a seta a lateral direita; (b) Imagem trimensional; (c) Perfil altimétrico médio da região destacada na imagem bidimensional.

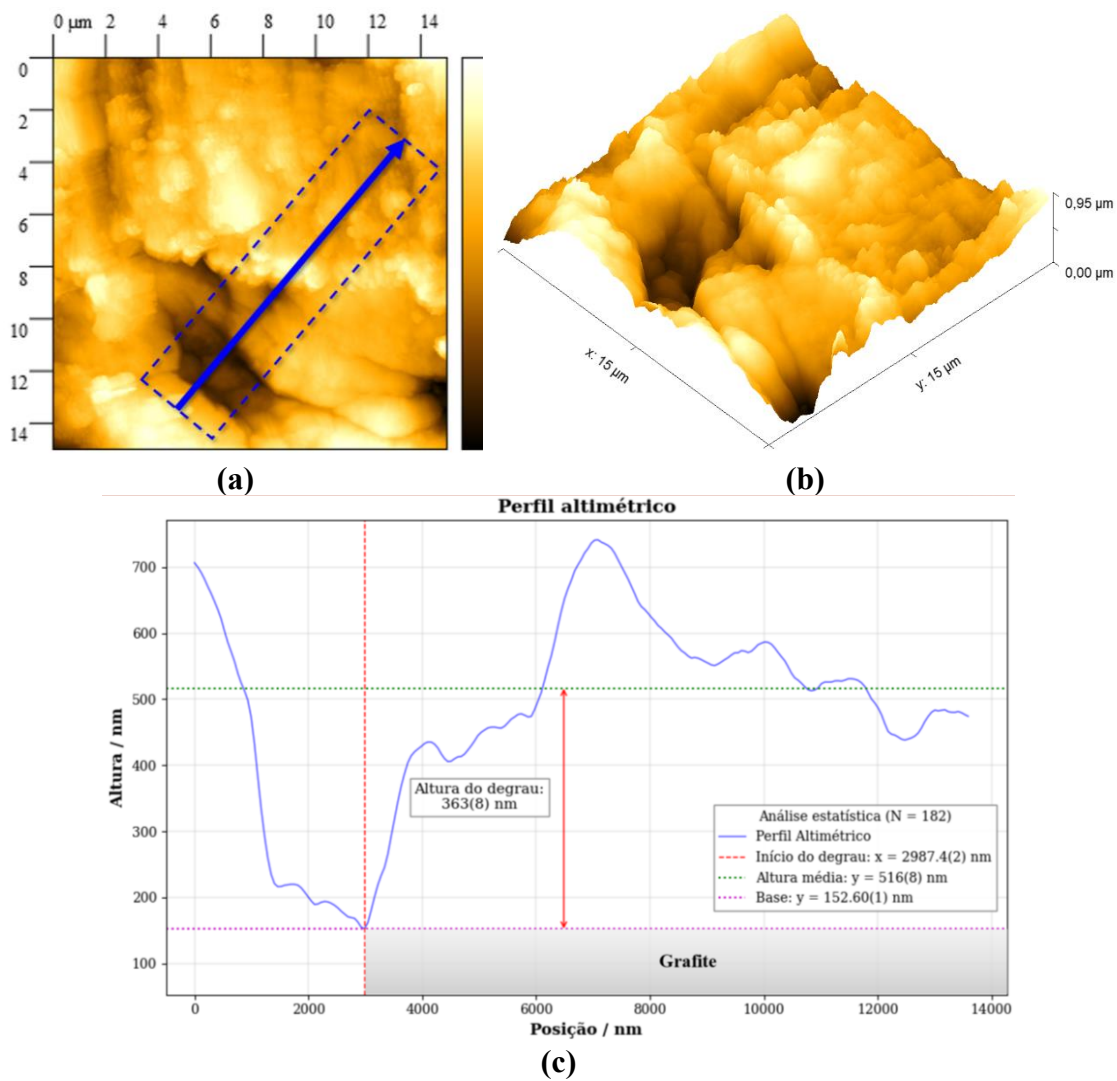

**Figura 14.** Figura de SPM obtida a apartir das análises de SPM: (a) Imagem bidimensional, o retângulo tracejado em azul indicaa área da qual foi extraído o perfil altimétrico médio e a seta indica o sentido correspondente no gráfico, sendo o início da seta a lateral esquerda do gráfico e a seta a lateral direita; (b) Imagem trimensional; (c) Perfil altimétrico médio da região destacada na imagem bidimensional.

A partir dos dados obtidos através da análise microscopia de varredura por sonda foi possível extrair perfis altimétricos de regiões de interesse, cujos dados são apresentados no Apêndice E. As regiões das quais extraiu-se os perfis estão destacadas com um retângulo tracejado em azul nas imagens bidimensionais (Figuras 11, 12 e 13 a) cujas setas indicam a lateral esquerda do gráfico (início da seta) e a lateral direita (final da seta). Para obter uma medida estatisticamente correta, obteve-se uma altura média das regiões consideradas para então calcular a diferença dessa altura média com o valor mais baixo do degrau, usado como valor de base do degrau. Nas Figuras (Figuras 11, 12 e 13 b) são indicados os dados que compõem o perfil altimétrico na cor azul; as bases dos degraus em linha tracejada na cor rosa; altura média do degrau em linha tracejada na cor verde; e o limite do degrau em linha tracejada na cor vermelha. Considerou-se a altura da camada de grafite depositada como sendo a diferença entre a linha "Base" e a "Altura média". Os valores dos degraus obtidos nas Figura 11, 12 e 13 são apresentados na Tabela 3.

**Tabela 3.** Medidas dos degraus das imagens de SPM.

| Figura | Degrau (nm) |
|--------|-------------|
| 11     | 500(17)     |
| 12     | 819(19)     |
| 13     | 363(8)      |
| Média  | 440(50)     |

A comparação entre os valores 456,5(34) nm (proposta do artigo) e 440(50) nm (SPM) pode ser feita analisando a sobreposição das incertezas e a significância da diferença entre as médias. Os intervalos de incerteza variam de 423,1 a 489,9 nm para os valores obtidos pela metodologia proposta e de 390 a 490 nm para as análises de SPM. Como esses intervalos se sobrepõem na faixa de 423,1 a 489,9 nm, as medições podem ser consideradas compatíveis dentro das incertezas experimentais associadas.

Além disso, a diferença entre os valores médios é de 16,5 nm, e a incerteza dessa diferença, obtida pela soma quadrática das incertezas individuais, é de aproximadamente 50 nm. O resultado 16(50) nm indica que a diferença não é estatisticamente significativa, pois o intervalo resultante inclui o valor zero — o limite inferior é cerca de –34 nm. Portanto, não há evidência estatística suficiente para afirmar que os valores medidos são diferentes. Isso sugere que os dois métodos de medição podem produzir resultados compatíveis dentro das incertezas experimentais, validando a proposta aqui relatada.

# CONSIDERAÇÕES FINAIS

O presente trabalho apresentou uma proposta experimental para a medição da resistividade do grafite 6B utilizando a Lei de Ohm e técnicas complementares como a microscopia de varredura por sonda (SPM). A metodologia adotada permitiu a determinação da espessura nanométrica dos traços de grafite utilizando equipamentos relativamente simples. Ademais, os valores obtidos se mostraram coerentes com os determinados por meio da SPM, já que as incertezas associadas às medições da espessura do grafite obtidas por ambas as metodologias possuem sobreposição dentro dos limites experimentais. Esse resultado valida a proposta do estudo, demonstrando que a metodologia proposta pode ser uma alternativa viável para estimar a espessura de camadas de grafite de forma acessível e eficiente. Além disso, a experiência foi concebida como uma atividade didática que pode ser aplicada em disciplinas de Eletricidade no Ensino Superior, proporcionando aos estudantes um experimento que integra conceitos teóricos e práticos. A relativa simplicidade da abordagem permite também que a proposta seja adaptada para o Ensino Médio, ampliando seu impacto educacional. Por fim, este estudo ressalta a importância de experimentos acessíveis para o ensino de conceitos fundamentais da Física, incentivando a aplicação de diferentes técnicas e ferramentas de análise na investigação de fenômenos elétricos e materiais condutores.

# REFERÊNCIAS

# APÊNDICE A

# DADOS DOS DIÂMETROS DO GRAFITE E CÁLCULO DA ÁREA DA SEÇÃO TRANSVERSAL

As medidas dos dois lados do lápis descascado são mostrados na tabela a seguir.

**Tabela 4.** Medidas dos diâmetros dos dois lados descascados do grafite

| Lado | Medida | Diâmetro - D (mm) |
|---|---|---|
| 1 | 1 | 2,81(1) |
| | 2 | 2,80(1) |
| | 3 | 2,80(1) |
| | 4 | 2,80(1) |
| | 5 | 2,80(1) |
| | 6 | 2,82(1) |
| | 7 | 2,83(1) |
| | 8 | 2,80(1) |
| | 9 | 2,79(1) |
| | 10 | 2,80(1) |
| 2 | 11 | 2,86(1) |
| | 12 | 2,79(1) |
| | 13 | 2,81(1) |
| | 14 | 2,79(1) |
| | 15 | 2,81(1) |
| | 16 | 2,81(1) |
| | 17 | 2,80(1) |
| | 18 | 2,84(1) |
| | 19 | 2,80(1) |
| | 20 | 2,79(1) |

Para o cálculo do erro da média dos diâmetros, calculou-se o desvio padrão do conjunto de dados ($\sigma$). A seguir são apresentados a média dos diâmetros ($\bar{D}$) e o valor de $\sigma$:

**Tabela 5.** Resultados estatísticos das medidas de diâmetros da Tabela 4.

| $\bar{D}$ | $\sigma$ | Undiade de medida | Observação |
|---|---|---|---|
| $2,8075 \cdot 10^{-3}$ | 0,0000175 | m | |
| $2,808 \cdot 10^{-3}$ | 0,000018 | m | Arredondado |

Para o cálculo do erro total, utilizou-se a equação da incerteza total ($u_{total}$) pois essa equação leva em consideração a incerteza do instrumento de medida ou da medida ($u_{med}$) e o desvio-padrão com relação a média do conjunto de dados. A equação de $u_{total}$ é apresentada a seguir:

$$u_{total} = \sqrt{(u_{ins})^2 + (\sigma_{media})^2} \qquad (19)$$

O valor de $u_{ins}$ trata-se da incerteza associada ao instrumento utilizado, enquanto que a $\sigma_{media}$ é a incerteza associada ao desvio padrão da média, calculada a partir dos dados utilizando seguinte equação:

$$\sigma_{media} = \frac{\sigma}{\sqrt{n}} \qquad (20)$$

Em que o desvio-padrão dos dados foi calculado a partir da seguinte equação:

$$\sigma = \sqrt{\frac{1}{n-1} \cdot \sum_{i=1}^{n}(x_i - \bar{x})^2} \qquad (21)$$

Em que $n$ é número de dados, $x_i$ é o dado de índice i e $\bar{x}$ é a média dos dados. Um resumo das grandezas estatísticas obtidas é demonstrado na Tabela 6.

**Tabela 6.** Grandezas estatísticas obtidas a partir das medidddas de diâmetros do grafite.

| | Grandeza estatística | Valor | Unidade de medida | Observação |
|---|---|---|---|---|
| $\sigma$ | Desvio padrão | 0,018 | mm | Equação 00 |
| $\sigma_{media}$ | Desvio padrão da média | 0,0040 | mm | Equação 00 |
| $u_{ins}$ | Incerteza instrumental | 0,01 | mm | |
| $u_{total}$ | Incerteza total | 0,0108 | mm | Equação 0 |
| | | 0,011 | mm | Arredondado |
| | | 0,000011 | m | Arredondado |

Desta forma, pode-se obter o seguinte valor $D = 2,808(11) \cdot 10^{-3}\ m$ para o valor do diâmetro, o qual foi utilizado nas etapas a seguir.

O cálculo da área foi aproximado para um círculo, sendo utilizada a seguinte equação:

$$A = \pi \cdot \frac{D^2}{4} \qquad (22)$$

Sendo necessário propagar os erros através da seguinte equação:

$$\sigma_A = \left|\frac{dA}{dD}\right| \cdot \sigma_A \qquad (23)$$

Sendo a derivada definida como:

$$\frac{dA}{dD} = \pi \cdot \frac{D}{2} \qquad (24)$$

Logo:

$$\sigma_A = \pi \cdot \frac{D}{2} \cdot \sigma_D \qquad (25)$$

O valor do diâmetro é $D = 2{,}81(1) \cdot 10^{-3}\ m$ e sua incerteza total é $\sigma_D = 0{,}01 \cdot 10^{-3}\ m$. Então, utilizando as Equações ( 22 ) e ( 25 ) e o valor de $\pi = 3{,}14159265358979\ (39)$ pode-se chegar ao valor da área de $A = 6{,}19(5) \cdot 10^{-6}\ m^2$.

# APÊNDICE B

# CÁLCULO DA RESISTIVIDADE DO GRAFITE

O gráfico mostrado na FIGURA foi plotado a partir dos dados mostrados na Tabela 0.

**Tabela 7.** Medidas de corrente elétrica (I) em função da tensão elétrica aplicada (V) entre as duas extremidades do grafite.

| Medida | V (V) | I (A) |
|---|---|---|
| 1* | 0,0131(1) | 0,00302(1) |
| 2 | 0,0887(1) | 0,02001(1) |
| 3 | 0,1572(1) | 0,03536(1) |
| 4 | 0,2332(1) | 0,05233(1) |
| 5 | 0,3093(1) | 0,06935(1) |
| 6 | 0,3776(1) | 0,08465(1) |
| 7 | 0,4537(1) | 0,10165(1) |
| 8 | 0,5218(1) | 0,11689(1) |
| 9 | 0,5978(1) | 0,13386(1) |
| 10 | 0,6735(1) | 0,15074(1) |
| 11 | 0,7417(1) | 0,16599(1) |
| 12 | 0,8172(1) | 0,18285(1) |
| 13 | 0,8928(1) | 0,19978(1) |
| 14 | 0,9605(1) | 0,21488(1) |
| 15 | 1,0357(1) | 0,23175(1) |
| 16 | 1,1106(1) | 0,24854(1) |
| 17 | 1,1779(1) | 0,26374(1) |
| 18 | 1,2524(1) | 0,28051(1) |
| 19 | 1,3270(1) | 0,29731(1) |
| 20 | 1,3934(1) | 0,31234(1) |
| 21 | 1,4675(1) | 0,32912(1) |
| 22 | 1,5333(1) | 0,34419(1) |
| 23 | 1,6069(1) | 0,36093(1) |
| 24 | 1,6792(1) | 0,37778(1) |
| 25 | 1,7440(1) | 0,39284(1) |
| 26 | 1,8163(1) | 0,40947(1) |
| 27 | 1,8873(1) | 0,42614(1) |
| 28 | 1,9515(1) | 0,44115(1) |
| 29 | 2,0215(1) | 0,45771(1) |
| 30 | 2,0918(1) | 0,47419(1) |
| 31 | 2,1537(1) | 0,48896(1) |
| 32 | 2,2218(1) | 0,50504(1) |
| 33 | 2,2890(1) | 0,52152(1) |
| 34 | 2,3489(1) | 0,53656(1) |
| 35 | 2,4139(1) | 0,55302(1) |
| 36 | 2,4719(1) | 0,56789(1) |
| 37 | 2,5340(1) | 0,58441(1) |
| 38 | 2,5965(1) | 0,60073(1) |
| 39 | 2,6512(1) | 0,61516(1) |
| 40 | 2,7109(1) | 0,63119(1) |
| 41 | 2,7682(1) | 0,6471(1) |
| 42 | 2,8178(1) | 0,66163(1) |
| 43 | 2,8726(1) | 0,67706(1) |
| 44 | 2,9260(1) | 0,69239(1) |
| 45 | 2,9707(1) | 0,70624(1) |
| 46 | 3,0212(1) | 0,72102(1) |
| 47 | 3,0648(1) | 0,73415(1) |
| 48 | 3,1100(1) | 0,74896(1) |
| 49 | 3,1542(1) | 0,76306(1) |
| 50 | 3,1938(1) | 0,77537(1) |
| 51 | 3,2338(1) | 0,78852(1) |

* Medida obtida quando a fonte está marcando 0 V. Mesmo que ela marque 0, ainda há uma corrente elétrica residual do sistema.

A resistividade ($\rho$) foi calculada a partir da Equação ( 14, sendo necessário a propagação de erro a apartir da seguinte equação:

$$\frac{\sigma_\rho}{\rho} = \sqrt{\left(\frac{\sigma_A}{A}\right)^2 + \left(\frac{\sigma_a}{a}\right)^2 + \left(\frac{\sigma_L}{L}\right)^2} \qquad (26)$$

Em que $\sigma_\rho$ é a incerteza da resistividade do grafite; A é a área da seção do grafite e $\sigma_A$ é sua incerteza associada (calculada no Apêndice A); a é o coeficiente angular obtido a partir do ajuste linear nos dados apresentados na Tabela 0 e $\sigma_a$ é a incerteza associada ao coeficiente angular; L é o comprimento do grafite e e $\sigma_L$ é a incerteza realcionada ao comprimento, no caso, a incerteza do isntrumento de medida utilziado. A Tabela 8 organiza os valores das grandezas envolvidas na Equação ( 14 e o resultado da propação de erro através da Equação ( 26.

**Tabela 8.** Valores das grandezas relacionadas ao cálculo da resistividade do grafite.

| Grandeza | Valor | Incerteza ($\sigma$) | Unidade de medida | Observação |
|---|---|---|---|---|
| A | $6,19 \cdot 10^{-6}$ | $0,05 \cdot 10^{-6}$ | m$^2$ | Apêndice A |
| a | $2,395 \cdot 10^{-1}$ | $0,016 \cdot 10^{-1}$ | $\Omega^{-1}$ | Obtido pelo ajuste linear dos dados da Tabela 7. |
| L | $1,3915 \cdot 10^{-1}$ | $0,0001 \cdot 10^{-1}$ | m | |
| $\rho$ | $1,85738 \cdot 10^{-4}$ | $0,019 \cdot 10^{-4}$ | $\Omega \cdot$m | Equação ( 14 ) e ( 26 |
| | $1,857 \cdot 10^{-4}$ | $0,019 \cdot 10^{-4}$ | $\Omega \cdot$m | Arredondado |

O valor de $\rho$ foi utilizado nas etapas posteriores.

# APÊNDICE C

# CÁLCULO DA LARGURA MÉDIA DOS TRECHOS DA CAMADA DE GRAFITE

As medidas das larguras da camada de grafite em diferentes trechos são apresentadas na Tabela 00.

**Tabela 9.** Medidas da largura dos trechos da camada de grafite.

| Trecho: 0 – 2 cm | | Trecho: 2 – 4 cm | | Trecho: 4 – 6 cm | | Trecho: 6 – 8 cm | | Trecho: 8 – 10 cm | |
|---|---|---|---|---|---|---|---|---|---|
| Medida | Largura (mm) | Medida | Largura (mm) | Medida | Largura (mm) | Medida | Largura (mm) | Medida | Largura (mm) |
| 1 | 2,52(1) | 51 | 2,35(1) | 101 | 2,54(1) | 151 | 2,43(1) | 201 | 2,66(1) |
| 2 | 2,57(1) | 52 | 2,39(1) | 102 | 2,52(1) | 152 | 2,56(1) | 202 | 2,37(1) |
| 3 | 2,59(1) | 53 | 2,42(1) | 103 | 2,57(1) | 153 | 2,56(1) | 203 | 2,70(1) |
| 4 | 2,33(1) | 54 | 2,43(1) | 104 | 2,47(1) | 154 | 2,57(1) | 204 | 2,42(1) |
| 5 | 2,48(1) | 55 | 2,19(1) | 105 | 2,52(1) | 155 | 2,50(1) | 205 | 2,49(1) |
| 6 | 2,52(1) | 56 | 2,26(1) | 106 | 2,59(1) | 156 | 2,31(1) | 206 | 2,44(1) |
| 7 | 2,40(1) | 57 | 2,33(1) | 107 | 2,56(1) | 157 | 2,56(1) | 207 | 2,25(1) |
| 8 | 2,28(1) | 58 | 2,63(1) | 108 | 2,43(1) | 158 | 2,55(1) | 208 | 1,78(1) |
| 9 | 2,41(1) | 59 | 2,21(1) | 109 | 2,45(1) | 159 | 2,55(1) | 209 | 2,66(1) |
| 10 | 2,35(1) | 60 | 2,54(1) | 110 | 2,54(1) | 160 | 2,47(1) | 210 | 2,53(1) |
| 11 | 2,32(1) | 61 | 2,40(1) | 111 | 2,56(1) | 161 | 2,61(1) | 211 | 2,38(1) |
| 12 | 2,44(1) | 62 | 2,40(1) | 112 | 2,56(1) | 162 | 1,98(1) | 212 | 2,54(1) |
| 13 | 2,38(1) | 63 | 2,46(1) | 113 | 2,57(1) | 163 | 2,28(1) | 213 | 2,23(1) |
| 14 | 2,48(1) | 64 | 2,46(1) | 114 | 2,45(1) | 164 | 2,24(1) | 214 | 2,39(1) |
| 15 | 2,34(1) | 65 | 2,25(1) | 115 | 2,52(1) | 165 | 2,33(1) | 215 | 2,35(1) |
| 16 | 2,10(1) | 66 | 2,25(1) | 116 | 2,50(1) | 166 | 2,47(1) | 216 | 1,56(1) |
| 17 | 2,39(1) | 67 | 2,39(1) | 117 | 2,28(1) | 167 | 2,54(1) | 217 | 2,13(1) |
| 18 | 2,49(1) | 68 | 2,19(1) | 118 | 2,39(1) | 168 | 2,08(1) | 218 | 1,99(1) |
| 19 | 2,40(1) | 69 | 2,09(1) | 119 | 2,48(1) | 169 | 2,51(1) | 219 | 2,51(1) |
| 20 | 2,45(1) | 70 | 1,76(1) | 120 | 2,48(1) | 170 | 2,56(1) | 220 | 2,38(1) |
| 21 | 2,57(1) | 71 | 2,11(1) | 121 | 2,59(1) | 171 | 2,58(1) | 221 | 2,34(1) |
| 22 | 2,56(1) | 72 | 2,23(1) | 122 | 2,26(1) | 172 | 2,50(1) | 222 | 2,37(1) |
| 23 | 2,49(1) | 73 | 1,94(1) | 123 | 2,49(1) | 173 | 2,58(1) | 223 | 2,39(1) |
| 24 | 2,43(1) | 74 | 1,77(1) | 124 | 2,21(1) | 174 | 2,45(1) | 224 | 2,55(1) |
| 25 | 2,10(1) | 75 | 2,22(1) | 125 | 2,47(1) | 175 | 2,64(1) | 225 | 2,27(1) |
| 26 | 2,42(1) | 76 | 2,15(1) | 126 | 2,26(1) | 176 | 2,43(1) | 226 | 2,21(1) |
| 27 | 2,38(1) | 77 | 1,95(1) | 127 | 2,51(1) | 177 | 2,63(1) | 227 | 2,00(1) |
| 28 | 2,49(1) | 78 | 2,08(1) | 128 | 2,25(1) | 178 | 2,62(1) | 228 | 2,47(1) |
| 29 | 2,36(1) | 79 | 2,17(1) | 129 | 2,44(1) | 179 | 2,44(1) | 229 | 2,30(1) |
| 30 | 2,29(1) | 80 | 2,13(1) | 130 | 2,36(1) | 180 | 2,05(1) | 230 | 2,38(1) |
| 31 | 2,09(1) | 81 | 2,39(1) | 131 | 2,64(1) | 181 | 1,59(1) | 231 | 2,30(1) |
| 32 | 2,43(1) | 82 | 2,43(1) | 132 | 2,42(1) | 182 | 1,81(1) | 232 | 2,47(1) |
| 33 | 2,53(1) | 83 | 2,25(1) | 133 | 2,57(1) | 183 | 2,58(1) | 233 | 2,40(1) |
| 34 | 2,49(1) | 84 | 2,09(1) | 134 | 2,18(1) | 184 | 2,42(1) | 234 | 2,17(1) |
| 35 | 2,59(1) | 85 | 2,09(1) | 135 | 2,57(1) | 185 | 2,42(1) | 235 | 2,55(1) |
| 36 | 2,35(1) | 86 | 2,21(1) | 136 | 2,46(1) | 186 | 2,54(1) | 236 | 2,11(1) |
| 37 | 2,12(1) | 87 | 2,14(1) | 137 | 2,43(1) | 187 | 2,49(1) | 237 | 2,31(1) |
| 38 | 2,54(1) | 88 | 2,19(1) | 138 | 2,51(1) | 188 | 2,12(1) | 238 | 1,47(1) |
| 39 | 2,50(1) | 89 | 1,94(1) | 139 | 2,59(1) | 189 | 2,61(1) | 239 | 1,91(1) |
| 40 | 2,44(1) | 90 | 2,36(1) | 140 | 2,61(1) | 190 | 2,54(1) | 240 | 2,67(1) |
| 41 | 2,13(1) | 91 | 1,80(1) | 141 | 2,61 | 191 | 2,56(1) | 241 | 1,93(1) |
| 42 | 2,39(1) | 92 | 1,73(1) | 142 | 2,42(1) | 192 | 2,44(1) | 242 | 2,60(1) |

| 43 | 2,35(1) | 93 | 1,85(1) | 143 | 2,03(1) | 193 | 2,49(1) | 243 | 1,96(1) |
| 44 | 2,50(1) | 94 | 1,95(1) | 144 | 2,54(1) | 194 | 2,61(1) | 244 | 2,42(1) |
| 45 | 2,27(1) | 95 | 2,53(1) | 145 | 2,50(1) | 195 | 2,42(1) | 245 | 2,35(1) |
| 46 | 2,36(1) | 96 | 2,46(1) | 146 | 2,11(1) | 196 | 2,42(1) | 246 | 1,88(1) |
| 47 | 1,85(1) | 97 | 2,30(1) | 147 | 2,39(1) | 197 | 2,55(1) | 247 | 2,61(1) |
| 48 | 2,37(1) | 98 | 2,01(1) | 148 | 2,53(1) | 198 | 2,63(1) | 248 | 2,50(1) |
| 49 | 2,44(1) | 99 | 1,93(1) | 149 | 2,09(1) | 199 | 2,21(1) | 249 | 2,44(1) |
| 50 | 2,46(1) | 100 | 1,80(1) | 150 | 2,31(1) | 200 | 2,38(1) | 250 | 2,02(1) |

A obtenção do valor da largura (b) de cada trecho foi calculada a partir da média aritmética dos dados correspondentes, bem como da incerteza total, conforme descrita na Equação ( 19 ). A Tabela 10 apresenta os valores obtidos.

**Tabela 10.** Grandezas obtidas de cada trecho da camada de grafite.

| Trecho | Grandeza | Valor | Unidade de Medida | Observação |
|---|---|---|---|---|
| 0 – 2 cm | n | 50 | | Quantidade de dados |
| | $\overline{b}$ | $2{,}39036 \cdot 10^{-3}$ | m | Intervalo das medidas da Tabela 9 usado: 1 – 50 |
| | $\sigma_{media}$ | $0{,}020967 \cdot 10^{-3}$ | m | Equação ( 20 ) |
| | $u_{ins}$ | $0{,}01 \cdot 10^{-3}$ | m | |
| | $u_{total}$ | $0{,}023 \cdot 10^{-3}$ | m | Equação ( 19 ) |
| | $b$ | $2{,}390(23) \cdot 10^{-3}$ | m | |
| 0 – 4 cm | n | 100 | | Intervalo das medidas da Tabela 9 usado: 1 – 100 |
| | $\overline{b}$ | $2{,}22907 \cdot 10^{-3}$ | m | |
| | $\sigma_{media}$ | $0{,}0215 \cdot 10^{-3}$ | m | Equação ( 20 ) |
| | $u_{ins}$ | $0{,}01 \cdot 10^{-3}$ | m | |
| | $u_{total}$ | $0{,}024 \cdot 10^{-3}$ | m | Equação ( 19 ) |
| | $b$ | $2{,}291(24) \cdot 10^{-3}$ | m | |
| 0 – 6 cm | n | 150 | | Intervalo das medidas da Tabela 9 usado: 1 – 150 |
| | $\overline{b}$ | $2{,}34239 \cdot 10^{-3}$ | m | |
| | $\sigma_{media}$ | $0{,}0169 \cdot 10^{-3}$ | m | Equação ( 20 ) |
| | $u_{ins}$ | $0{,}01 \cdot 10^{-3}$ | m | |
| | $u_{total}$ | $0{,}0196 \cdot 10^{-3}$ | m | Equação ( 19 ) |
| | $b$ | $2{,}342(20) \cdot 10^{-3}$ | m | |
| 0 – 8 cm | n | 200 | | Intervalo das medidas da Tabela 9 usado: 1 – 200 |
| | $\overline{b}$ | $2{,}36366 \cdot 10^{-3}$ | m | |
| | $\sigma_{media}$ | $0{,}01505 \cdot 10^{-3}$ | m | Equação ( 20 ) |
| | $u_{ins}$ | $0{,}01 \cdot 10^{-3}$ | m | |
| | $u_{total}$ | $0{,}01807 \cdot 10^{-3}$ | m | Equação ( 19 ) |
| | $b$ | $2{,}364(18) \cdot 10^{-3}$ | m | |
| 0 – 10 cm | n | 250 | | Intervalo das medidas da Tabela 9 usado: 1 – 250 |
| | $\overline{b}$ | $2{,}35145 \cdot 10^{-3}$ | m | |
| | $\sigma_{media}$ | $0{,}01439 \cdot 10^{-3}$ | m | Equação ( 20 ) |
| | $u_{ins}$ | $0{,}01 \cdot 10^{-3}$ | m | |

| | | | |
|---|---|---|---|
| $u_{total}$ | $0{,}01752 \cdot 10^{-3}$ | m | Equação ( 19 ) |
| $b$ | $2{,}351(18) \cdot 10^{-3}$ | m | |

# APÊNDICE D

# CÁLCULO DA ALTURA DA CAMADA DE GRAFITE DEPOSITADA SOBRE PAPEL VEGETAL

**Tabela 11.** Medidas de corrente elétrica (I) em função da tensão elétrica aplicada (V) do trecho 0 – 2 cm.

| Medida | V (V) | I (A) |
|--------|-------|-------|
| 1* | 0,0175(1) | 0,000004(1) |
| 2 | 0,1212(1) | 0,000037(1) |
| 3 | 0,2142(1) | 0,000066(1) |
| 4 | 0,3181(1) | 0,000098(1) |
| 5 | 0,4215(1) | 0,00013(1) |
| 6 | 0,5150(1) | 0,000159(1) |
| 7 | 0,6184(1) | 0,000191(1) |
| 8 | 0,7119(1) | 0,00022(1) |
| 9 | 0,8154(1) | 0,000252(1) |
| 10 | 0,9193(1) | 0,000286(1) |
| 11 | 1,0124(1) | 0,000314(1) |
| 12 | 1,1161(1) | 0,000347(1) |
| 13 | 1,2199(1) | 0,000379(1) |
| 14 | 1,3131(1) | 0,000408(1) |
| 15 | 1,4169(1) | 0,000441(1) |
| 16 | 1,5203(1) | 0,000472(1) |
| 17 | 1,6139(1) | 0,000501(1) |
| 18 | 1,7173(1) | 0,000534(1) |
| 19 | 1,8212(1) | 0,000567(1) |
| 20 | 1,9143(1) | 0,000596(1) |
| 21 | 2,0181(1) | 0,000628(1) |
| 22 | 2,1114(1) | 0,000658(1) |
| 23 | 2,2147(1) | 0,000691(1) |
| 24 | 2,3187(1) | 0,000724(1) |
| 25 | 2,4117(1) | 0,000753(1) |
| 26 | 2,5156(1) | 0,000785(1) |
| 27 | 2,6190(1) | 0,000817(1) |
| 28 | 2,7127(1) | 0,000847(1) |
| 29 | 2,8161(1) | 0,00088(1) |
| 30 | 2,9200(1) | 0,000912(1) |
| 31 | 3,0130(1) | 0,000942(1) |
| 32 | 3,1170(1) | 0,000974(1) |
| 33 | 3,2206(1) | 0,001007(1) |
| 34 | 3,3136(1) | 0,001037(1) |
| 35 | 3,4175(1) | 0,001069(1) |
| 36 | 3,5105(1) | 0,0011(1) |
| 37 | 3,6146(1) | 0,001134(1) |
| 38 | 3,7178(1) | 0,001166(1) |
| 39 | 3,8115(1) | 0,001196(1) |
| 40 | 3,9148(1) | 0,001229(1) |
| 41 | 4,0188(1) | 0,001263(1) |
| 42 | 4,1117(1) | 0,001293(1) |
| 43 | 4,2158(1) | 0,001326(1) |
| 44 | 4,3190(1) | 0,001358(1) |
| 45 | 4,4128(1) | 0,001389(1) |
| 46 | 4,5160(1) | 0,001422(1) |
| 47 | 4,6098(1) | 0,00145(1) |
| 48 | 4,7136(1) | 0,001469(1) |
| 49 | 4,8168(1) | 0,00151(1) |
| 50 | 4,9102(1) | 0,001543(1) |
| 51 | 5,0145(1) | 0,001577(1) |

**Tabela 12.** Valores das grandezas relacionadas ao cálculo da altura da camada de grafite depositada sobre a folha vegetal do trecho 0 – 2 cm.

| Grandeza | Valor | Incerteza ($\sigma$) | Unidade de medida | Observação |
|---|---|---|---|---|
| $\rho$ | $1,857 \cdot 10^{-4}$ | $0,019 \cdot 10^{-4}$ | $\Omega \cdot m$ | Apêndice B - Tabela 8 |
| L | $2,00 \cdot 10^{-2}$ | $0,05 \cdot 10^{-2}$ | m | |
| b | $2,390 \cdot 10^{-3}$ | $0,023 \cdot 10^{-3}$ | m | Apêndice C - Tabela 12 |
| $a$ | $3,1478 \cdot 10^{-4}$ | $0,0026 \cdot 10^{-4}$ | $\Omega^{-1}$ | Obtido pelo ajuste linear dos dados da Tabela 11. |
| h | $4,89262 \cdot 10^{-7}$ | $0,14035 \cdot 10^{-7}$ | m | Equação ( 10 ) |
| | $4,89 \cdot 10^{-7}$ | $0,14 \cdot 10^{-7}$ | m | Arredondado |
| | 489 | 14 | nm | |

**Tabela 13.** Medidas de corrente elétrica (I) em função da tensão elétrica aplicada (V) do trecho 0 – 4 cm.

| Medida | V (V) | I (A) |
|---|---|---|
| 1* | 0,0177(1) | 0,002(1) |
| 2 | 0,1210(1) | 0,020(1) |
| 3 | 0,2145(1) | 0,036(1) |
| 4 | 0,3180(1) | 0,054(1) |
| 5 | 0,4218(1) | 0,073(1) |
| 6 | 0,5150(1) | 0,089(1) |
| 7 | 0,6188(1) | 0,108(1) |
| 8 | 0,7120(1) | 0,125(1) |
| 9 | 0,8159(1) | 0,143(1) |
| 10 | 0,9193(1) | 0,161(1) |
| 11 | 1,0129(1) | 0,177(1) |
| 12 | 1,1164(1) | 0,196(1) |
| 13 | 1,2201(1) | 0,214(1) |
| 14 | 1,3133(1) | 0,230(1) |
| 15 | 1,4173(1) | 0,249(1) |
| 16 | 1,5209(1) | 0,267(1) |
| 17 | 1,6140(1) | 0,284(1) |
| 18 | 1,7178(1) | 0,302(1) |
| 19 | 1,8213(1) | 0,320(1) |
| 20 | 1,9149(1) | 0,336(1) |
| 21 | 2,0182(1) | 0,355(1) |
| 22 | 2,1119(1) | 0,371(1) |
| 23 | 2,2152(1) | 0,390(1) |
| 24 | 2,3192(1) | 0,408(1) |
| 25 | 2,4122(1) | 0,425(1) |
| 26 | 2,5162(1) | 0,443(1) |
| 27 | 2,6196(1) | 0,462(1) |
| 28 | 2,7132(1) | 0,478(1) |
| 29 | 2,8166(1) | 0,496(1) |
| 30 | 2,9206(1) | 0,514(1) |
| 31 | 3,0135(1) | 0,531(1) |
| 32 | 3,1176(1) | 0,549(1) |
| 33 | 3,2209(1) | 0,568(1) |
| 34 | 3,3146(1) | 0,584(1) |
| 35 | 3,4183(1) | 0,603(1) |
| 36 | 3,5113(1) | 0,619(1) |
| 37 | 3,6154(1) | 0,638(1) |
| 38 | 3,7186(1) | 0,656(1) |
| 39 | 3,8122(1) | 0,673(1) |
| 40 | 3,9157(1) | 0,692(1) |
| 41 | 4,0196(1) | 0,710(1) |
| 42 | 4,1127(1) | 0,726(1) |
| 43 | 4,2168(1) | 0,744(1) |
| 44 | 4,3200(1) | 0,763(1) |
| 45 | 4,4137(1) | 0,780(1) |
| 46 | 4,5171(1) | 0,799(1) |
| 47 | 4,6108(1) | 0,815(1) |
| 48 | 4,7142(1) | 0,833(1) |
| 49 | 4,8182(1) | 0,852(1) |
| 50 | 4,9115(1) | 0,869(1) |
| 51 | 5,0150(1) | 0,887(1) |

**Tabela 14.** Valores das grandezas relacionadas ao cálculo da altura da camada de grafite depositada sobre a folha vegetal do trecho 0 – 4 cm.

| Grandeza | Valor | Incerteza ($\sigma$) | Unidade de medida | Observação |
|---|---|---|---|---|
| $\rho$ | $1{,}857 \cdot 10^{-4}$ | $0{,}019 \cdot 10^{-4}$ | $\Omega \cdot m$ | Apêndice B - Tabela 8 |
| $L$ | $4{,}00 \cdot 10^{-2}$ | $0{,}05 \cdot 10^{-2}$ | m | |
| b | $2{,}291 \cdot 10^{-3}$ | $0{,}024 \cdot 10^{-3}$ | m | Apêndice C - Tabela 12 |
| $a$ | $1{,}7711 \cdot 10^{-4}$ | $0{,}00006 \cdot 10^{-4}$ | $\Omega^{-1}$ | Obtido pelo ajuste linear dos dados da Tabela 13. |
| h | 5,7435 · 10⁻⁷ | 0,11059 · 10⁻⁷ | m | Equação ( 10 ) |
| | 5,74 · 10⁻⁷ | 0,11 · 10⁻⁷ | m | Arredondado |
| | 574 | 11 | nm | |

**Tabela 15.** Medidas de corrente elétrica (I) em função da tensão elétrica aplicada (V) do trecho 0 – 6 cm.

| Medida | V (V) | I (A) |
|---|---|---|
| 1* | 0,0177(1) | 0,000001(1) |
| 2 | 0,1212(1) | 0,000011(1) |
| 3 | 0,2147(1) | 0,000021(1) |
| 4 | 0,3184(1) | 0,000031(1) |
| 5 | 0,4221(1) | 0,00004(1) |
| 6 | 0,5154(1) | 0,00005(1) |
| 7 | 0,6191(1) | 0,000061(1) |
| 8 | 0,7123(1) | 0,00007(1) |
| 9 | 0,8162(1) | 0,000081(1) |
| 10 | 0,9197(1) | 0,000092(1) |
| 11 | 1,0132(1) | 0,0001(1) |
| 12 | 1,1168(1) | 0,000111(1) |
| 13 | 1,2205(1) | 0,000122(1) |
| 14 | 1,3138(1) | 0,000131(1) |
| 15 | 1,4176(1) | 0,000142(1) |
| 16 | 1,5212(1) | 0,000152(1) |
| 17 | 1,6148(1) | 0,000163(1) |
| 18 | 1,7182(1) | 0,000173(1) |
| 19 | 1,8221(1) | 0,000184(1) |
| 20 | 1,9153(1) | 0,000193(1) |
| 21 | 2,0192(1) | 0,000204(1) |
| 22 | 2,1122(1) | 0,000216(1) |
| 23 | 2,2163(1) | 0,000226(1) |
| 24 | 2,3197(1) | 0,000237(1) |
| 25 | 2,4133(1) | 0,000247(1) |
| 26 | 2,5166(1) | 0,000258(1) |
| 27 | 2,6207(1) | 0,00027(1) |
| 28 | 2,7138(1) | 0,00028(1) |
| 29 | 2,8178(1) | 0,000291(1) |
| 30 | 2,9211(1) | 0,000302(1) |
| 31 | 3,0148(1) | 0,000312(1) |
| 32 | 3,1181(1) | 0,000323(1) |
| 33 | 3,2222(1) | 0,000334(1) |
| 34 | 3,3152(1) | 0,000345(1) |
| 35 | 3,4193(1) | 0,000356(1) |
| 36 | 3,5123(1) | 0,000367(1) |
| 37 | 3,6164(1) | 0,000378(1) |
| 38 | 3,7195(1) | 0,000389(1) |
| 39 | 3,8134(1) | 0,000399(1) |
| 40 | 3,9167(1) | 0,000409(1) |
| 41 | 4,0207(1) | 0,00042(1) |
| 42 | 4,1137(1) | 0,000431(1) |
| 43 | 4,2178(1) | 0,000442(1) |
| 44 | 4,3212(1) | 0,000453(1) |
| 45 | 4,4149(1) | 0,000464(1) |
| 46 | 4,5183(1) | 0,000476(1) |
| 47 | 4,6121(1) | 0,000485(1) |
| 48 | 4,7154(1) | 0,000498(1) |
| 49 | 4,8196(1) | 0,000509(1) |
| 50 | 4,9125(1) | 0,00052(1) |
| 51 | 5,0167(1) | 0,000532(1) |

**Tabela 16.** Valores das grandezas relacionadas ao cálculo da altura da camada de grafite depositada sobre a folha vegetal do trecho 0 – 6 cm.

| Grandeza | Valor | Incerteza ($\sigma$) | Unidade de medida | Observação |
|---|---|---|---|---|
| $\rho$ | $1,857 \cdot 10^{-4}$ | $0,019 \cdot 10^{-4}$ | $\Omega \cdot m$ | Apêndice B - Tabela 8 |
| $L$ | $6,00 \cdot 10^{-2}$ | $0,05 \cdot 10^{-2}$ | m | |
| b | $2,342 \cdot 10^{-3}$ | $0,020 \cdot 10^{-3}$ | m | Apêndice C - Tabela 12 |
| $a$ | $1,0637 \cdot 10^{-4}$ | $0,0025 \cdot 10^{-5}$ | $\Omega^{-1}$ | Obtido pelo ajuste linear dos dados da Tabela 11. |
| h | $5,06157 \cdot 10^{-7}$ | $0,08043 \cdot 10^{-7}$ | m | Equação ( 10 ) |
| | $5,06 \cdot 10^{-7}$ | $0,08 \cdot 10^{-7}$ | m | Arredondado |
| | 506 | 8 | nm | |

**Tabela 17.** Medidas de corrente elétrica (I) em função da tensão elétrica aplicada (V) do trecho 0 – 8 cm.

| Medida | V (V) | I (A) |
|---|---|---|
| 1* | 0,0174(1) | 0,000001(1) |
| 2 | 0,1213(1) | 0,000007(1) |
| 3 | 0,2145(1) | 0,000013(1) |
| 4 | 0,3184(1) | 0,000019(1) |
| 5 | 0,4220(1) | 0,000026(1) |
| 6 | 0,5155(1) | 0,000031(1) |
| 7 | 0,6189(1) | 0,000038(1) |
| 8 | 0,7125(1) | 0,000043(1) |
| 9 | 0,8160(1) | 0,000049(1) |
| 10 | 0,9199(1) | 0,000056(1) |
| 11 | 1,0130(1) | 0,000062(1) |
| 12 | 1,1169(1) | 0,000068(1) |
| 13 | 1,2204(1) | 0,000075(1) |
| 14 | 1,3140(1) | 0,000081(1) |
| 15 | 1,4174(1) | 0,000087(1) |
| 16 | 1,5214(1) | 0,000093(1) |
| 17 | 1,6145(1) | 0,000099(1) |
| 18 | 1,7184(1) | 0,000106(1) |
| 19 | 1,8218(1) | 0,000112(1) |
| 20 | 1,9155(1) | 0,000118(1) |
| 21 | 2,0189(1) | 0,000125(1) |
| 22 | 2,1125(1) | 0,00013(1) |
| 23 | 2,2159(1) | 0,000137(1) |
| 24 | 2,3199(1) | 0,000144(1) |
| 25 | 2,4129(1) | 0,000149(1) |
| 26 | 2,5169(1) | 0,000156(1) |
| 27 | 2,6204(1) | 0,000163(1) |
| 28 | 2,7140(1) | 0,000168(1) |
| 29 | 2,8174(1) | 0,000175(1) |
| 30 | 2,9215(1) | 0,000182(1) |
| 31 | 3,0148(1) | 0,000187(1) |
| 32 | 3,1182(1) | 0,000194(1) |
| 33 | 3,2223(1) | 0,000200(1) |
| 34 | 3,3153(1) | 0,000207(1) |
| 35 | 3,4193(1) | 0,000214(1) |
| 36 | 3,5124(1) | 0,000219(1) |
| 37 | 3,6164(1) | 0,000226(1) |
| 38 | 3,7197(1) | 0,000233(1) |
| 39 | 3,8134(1) | 0,000239(1) |
| 40 | 3,9168(1) | 0,000246(1) |
| 41 | 4,0209(1) | 0,000253(1) |
| 42 | 4,1139(1) | 0,000259(1) |
| 43 | 4,2180(1) | 0,000265(1) |
| 44 | 4,3213(1) | 0,000272(1) |
| 45 | 4,4151(1) | 0,000279(1) |
| 46 | 4,5184(1) | 0,000286(1) |
| 47 | 4,6122(1) | 0,000291(1) |
| 48 | 4,7155(1) | 0,000298(1) |
| 49 | 4,8198(1) | 0,000305(1) |
| 50 | 4,9127(1) | 0,000311(1) |
| 51 | 5,0170(1) | 0,000318(1) |

**Tabela 18.** Valores das grandezas relacionadas ao cálculo da altura da camada de grafite depositada sobre a folha vegetal do trecho 0 – 8 cm.

| Grandeza | Valor | Incerteza ($\sigma$) | Unidade de medida | Observação |
|---|---|---|---|---|
| $\rho$ | $1,857 \cdot 10^{-4}$ | $0,019 \cdot 10^{-4}$ | $\Omega \cdot m$ | Apêndice B - Tabela 8 |
| L | $8,00 \cdot 10^{-2}$ | $0,05 \cdot 10^{-2}$ | m | |
| b | $2,364 \cdot 10^{-3}$ | $0,018 \cdot 10^{-3}$ | m | Apêndice C - Tabela 12 |
| a | $6,343 \cdot 10^{-5}$ | $0,011 \cdot 10^{-5}$ | $\Omega^{-1}$ | Obtido pelo ajuste linear dos dados da Tabela 17. |
| h | $3,9869 \cdot 10^{-7}$ | $0,05704 \cdot 10^{-7}$ | m | Equação ( 10 ) |
| | $3,99 \cdot 10^{-7}$ | $0,06 \cdot 10^{-7}$ | m | Arredondado |

| | 399 | 6 | nm |
|---|---|---|---|

**Tabela 19.** Medidas de corrente elétrica (I) em função da tensão elétrica aplicada (V) do trecho 0 – 10 cm.

| Medida | V (V) | I (A) |
|---|---|---|
| 1* | 0,0175(1) | 0,000000(1) |
| 2 | 0,1213(1) | 0,000005(1) |
| 3 | 0,2146(1) | 0,000011(1) |
| 4 | 0,3184(1) | 0,000017(1) |
| 5 | 0,4218(1) | 0,000022(1) |
| 6 | 0,5154(1) | 0,000027(1) |
| 7 | 0,6189(1) | 0,000034(1) |
| 8 | 0,7124(1) | 0,000038(1) |
| 9 | 0,8161(1) | 0,000044(1) |
| 10 | 0,9199(1) | 0,00005(1) |
| 11 | 1,0131(1) | 0,000055(1) |
| 12 | 1,1169(1) | 0,000061(1) |
| 13 | 1,2204(1) | 0,000068(1) |
| 14 | 1,3140(1) | 0,000072(1) |
| 15 | 1,4174(1) | 0,000078(1) |
| 16 | 1,5214(1) | 0,000084(1) |
| 17 | 1,6146(1) | 0,000089(1) |
| 18 | 1,7184(1) | 0,000095(1) |
| 19 | 1,8221(1) | 0,000101(1) |
| 20 | 1,9153(1) | 0,000107(1) |
| 21 | 2,0191(1) | 0,000112(1) |
| 22 | 2,1122(1) | 0,000117(1) |
| 23 | 2,2163(1) | 0,000122(1) |
| 24 | 2,3197(1) | 0,000128(1) |
| 25 | 2,4133(1) | 0,000133(1) |
| 26 | 2,5167(1) | 0,000139(1) |
| 27 | 2,6207(1) | 0,000145(1) |
| 28 | 2,7139(1) | 0,000152(1) |
| 29 | 2,8179(1) | 0,000157(1) |
| 30 | 2,9212(1) | 0,000163(1) |
| 31 | 3,0149(1) | 0,00017(1) |
| 32 | 3,1183(1) | 0,000175(1) |
| 33 | 3,2223(1) | 0,000181(1) |
| 34 | 3,3154(1) | 0,000187(1) |
| 35 | 3,4194(1) | 0,000192(1) |
| 36 | 3,5125(1) | 0,000198(1) |
| 37 | 3,6164(1) | 0,000204(1) |
| 38 | 3,7199(1) | 0,000211(1) |
| 39 | 3,8135(1) | 0,000215(1) |
| 40 | 3,9168(1) | 0,00022(1) |
| 41 | 4,0209(1) | 0,000226(1) |
| 42 | 4,1139(1) | 0,000232(1) |
| 43 | 4,2180(1) | 0,000238(1) |
| 44 | 4,3214(1) | 0,000245(1) |
| 45 | 4,4152(1) | 0,00025(1) |
| 46 | 4,5185(1) | 0,000256(1) |
| 47 | 4,6123(1) | 0,000261(1) |
| 48 | 4,7156(1) | 0,000267(1) |
| 49 | 4,8199(1) | 0,000275(1) |
| 50 | 4,9128(1) | 0,000281(1) |
| 51 | 5,0170(1) | 0,000287(1) |

**Tabela 20.** Valores das grandezas relacionadas ao cálculo da altura da camada de grafite depositada sobre a folha vegetal do trecho 0 – 10 cm.

| Grandeza | Valor | Incerteza ($\sigma$) | Unidade de medida | Observação |
|---|---|---|---|---|
| $\rho$ | $1{,}857 \cdot 10^{-4}$ | $0{,}019 \cdot 10^{-4}$ | $\Omega \cdot m$ | Apêndice B - Tabela 8 |
| $L$ | $1{,}00 \cdot 10^{-1}$ | $0{,}005 \cdot 10^{-1}$ | m | |
| b | $2{,}351 \cdot 10^{-3}$ | $0{,}018 \cdot 10^{-3}$ | m | Apêndice C - Tabela 12 |
| $a$ | $5{,}718 \cdot 10^{-5}$ | $0{,}011 \cdot 10^{-5}$ | $\Omega^{-1}$ | Obtido pelo ajuste linear dos dados da Tabela 11. |
| h | $4{,}51745 \cdot 10^{-7}$ | $0{,}06258 \cdot 10^{-7}$ | m | Equação ( 10 ) |
| | $4{,}52 \cdot 10^{-7}$ | $0{,}06 \cdot 10^{-7}$ | m | Arredondado |
| | 452 | 6 | nm | |

Com os valores da altura da camada sobre o papel vegetal considerando diferentes trechos obtidos, pode-se calcular uma média ponderada conforme as diferentes incertezas associadas, a partir da seguinte equação:

$$\bar{x} = \frac{\sum\left(\frac{x_i}{\sigma_i^2}\right)}{\sum\left(\frac{1}{\sigma_i^2}\right)} \quad (27)$$

Em que $x_i$ é cada medida, no caso as diferentes alturas de camada de grafite obtidas; e $\sigma_i$ é a incerteza de cada medida, respectivamente. Para o cálculo da incerteza associado a essa média ponderada ($\sigma_{\bar{x}}$), utiliza-se a seguinte Equação:

$$\sigma_{\bar{x}} = \sqrt{\frac{1}{\sum\left(\frac{1}{\sigma_i^2}\right)}} \quad (28)$$

A partir dos dados da Tabela 2, pode-se calcular o valor da média apresentado no final da tabela, sendo de 390,0(28).

# APÊNDICE E

# MÉTODO DE PROCESSAMENTO DAS IMAGENS DE AFM E TRATAMENTO ESTATÍSTICOS DOS PERFIS ALTIMÉTRICOS

As imagens foram tratadas utilizando o *software* livre Gwyddion, versão 2.68, realizando as etapas de tratamentos estatísivtos das imagens apresentadas na Tabela 21.

**Tabela 21.** Relatório de tratamento das iamgens de AFM.

| Etapa | Comando | Objetivo |
|---|---|---|
| 1 | Remove polynomial background (2º grau) | Corrigir a inclinação ou curvatura da imagem ao subtrair um fundo ajustado por um polinômio de 2º grau |
| 2 | Correct horizontal scars (strokes) | Remover linhas horizontais indesejadas |

Após o processamento das imagens, foram extraídos os perfis das regiões destacadas por retângulos azuis nas Figuras Figura *12*, Figura **13** e Figura **14** a. Como os perfis representam uma área, foi necessário aplicar uma interpolação linear entre os dados correspondentes ao mesmo valor de "x", com o objetivo de obter a altimetria média. Os dados dos perfis são apresentados nas Tabelas Tabela *24*, Tabela *23* e Tabela *22*. A incerteza associada

**Tabela 22.** Dados utilizados para plotar o gráfico da Figura 14 c.

| Medida | Posição x (nm) | Altura z (nm) | Medida | Posição x (nm) | Altura z (nm) | Medida | Posição x (nm) | Altura z (nm) |
|---|---|---|---|---|---|---|---|---|
| 1 | 0,0(1) | 637,50(1) | 87 | 5039,1(1) | 1307,29(1) | 173 | 10078,1(1) | 891,11(1) |
| 2 | 58,6(1) | 652,10(1) | 88 | 5097,7(1) | 1304,27(1) | 174 | 10136,7(1) | 903,22(1) |
| 3 | 117,2(1) | 666,53(1) | 89 | 5156,3(1) | 1299,43(1) | 175 | 10195,3(1) | 918,13(1) |
| 4 | 175,8(1) | 691,65(1) | 90 | 5214,8(1) | 1296,26(1) | 176 | 10253,9(1) | 931,53(1) |
| 5 | 234,4(1) | 718,56(1) | 91 | 5273,4(1) | 1294,64(1) | 177 | 10312,5(1) | 951,36(1) |
| 6 | 293,0(1) | 739,03(1) | 92 | 5332,0(1) | 1293,19(1) | 178 | 10371,1(1) | 962,67(1) |
| 7 | 351,6(1) | 754,87(1) | 93 | 5390,6(1) | 1286,83(1) | 179 | 10429,7(1) | 966,74(1) |
| 8 | 410,2(1) | 769,61(1) | 94 | 5449,2(1) | 1278,44(1) | 180 | 10488,3(1) | 971,13(1) |
| 9 | 468,8(1) | 783,62(1) | 95 | 5507,8(1) | 1276,53(1) | 181 | 10546,9(1) | 980,57(1) |
| 10 | 527,3(1) | 799,77(1) | 96 | 5566,4(1) | 1273,54(1) | 182 | 10605,5(1) | 996,50(1) |
| 11 | 585,9(1) | 813,85(1) | 97 | 5625,0(1) | 1268,68(1) | 183 | 10664,1(1) | 1005,25(1) |
| 12 | 644,5(1) | 825,97(1) | 98 | 5683,6(1) | 1263,47(1) | 184 | 10722,7(1) | 1012,32(1) |
| 13 | 703,1(1) | 840,92(1) | 99 | 5742,2(1) | 1258,89(1) | 185 | 10781,3(1) | 1024,12(1) |
| 14 | 761,7(1) | 855,40(1) | 100 | 5800,8(1) | 1255,86(1) | 186 | 10839,8(1) | 1030,49(1) |
| 15 | 820,3(1) | 870,63(1) | 101 | 5859,4(1) | 1251,85(1) | 187 | 10898,4(1) | 1038,90(1) |
| 16 | 878,9(1) | 886,99(1) | 102 | 5918,0(1) | 1247,07(1) | 188 | 10957,0(1) | 1046,90(1) |
| 17 | 937,5(1) | 909,04(1) | 103 | 5976,6(1) | 1242,63(1) | 189 | 11015,6(1) | 1057,56(1) |
| 18 | 996,1(1) | 936,16(1) | 104 | 6035,2(1) | 1233,70(1) | 190 | 11074,2(1) | 1072,18(1) |

| | | | | | | | |
|---|---|---|---|---|---|---|---|
| 19 | 1054,7(1) | 955,96(1) | 105 | 6093,8(1) | 1219,62(1) | 191 | 11132,8(1) | 1086,60(1) |
| 20 | 1113,3(1) | 972,40(1) | 106 | 6152,3(1) | 1208,24(1) | 192 | 11191,4(1) | 1100,21(1) |
| 21 | 1171,9(1) | 987,50(1) | 107 | 6210,9(1) | 1201,15(1) | 193 | 11250,0(1) | 1112,81(1) |
| 22 | 1230,5(1) | 999,52(1) | 108 | 6269,5(1) | 1193,80(1) | 194 | 11308,6(1) | 1118,79(1) |
| 23 | 1289,1(1) | 1005,70(1) | 109 | 6328,1(1) | 1187,35(1) | 195 | 11367,2(1) | 1121,26(1) |
| 24 | 1347,7(1) | 1012,22(1) | 110 | 6386,7(1) | 1178,20(1) | 196 | 11425,8(1) | 1132,75(1) |
| 25 | 1406,3(1) | 1018,47(1) | 111 | 6445,3(1) | 1171,51(1) | 197 | 11484,4(1) | 1149,34(1) |
| 26 | 1464,8(1) | 1023,82(1) | 112 | 6503,9(1) | 1165,35(1) | 198 | 11543,0(1) | 1162,81(1) |
| 27 | 1523,4(1) | 1033,86(1) | 113 | 6562,5(1) | 1157,56(1) | 199 | 11601,6(1) | 1165,73(1) |
| 28 | 1582,0(1) | 1045,98(1) | 114 | 6621,1(1) | 1145,80(1) | 200 | 11660,2(1) | 1167,13(1) |
| 29 | 1640,6(1) | 1056,09(1) | 115 | 6679,7(1) | 1132,69(1) | 201 | 11718,8(1) | 1167,99(1) |
| 30 | 1699,2(1) | 1064,57(1) | 116 | 6738,3(1) | 1116,32(1) | 202 | 11777,3(1) | 1167,03(1) |
| 31 | 1757,8(1) | 1071,90(1) | 117 | 6796,9(1) | 1095,99(1) | 203 | 11835,9(1) | 1164,82(1) |
| 32 | 1816,4(1) | 1076,36(1) | 118 | 6855,5(1) | 1075,28(1) | 204 | 11894,5(1) | 1160,99(1) |
| 33 | 1875,0(1) | 1081,86(1) | 119 | 6914,1(1) | 1053,05(1) | 205 | 11953,1(1) | 1154,00(1) |
| 34 | 1933,6(1) | 1088,97(1) | 120 | 6972,7(1) | 1028,41(1) | 206 | 12011,7(1) | 1148,23(1) |
| 35 | 1992,2(1) | 1096,71(1) | 121 | 7031,3(1) | 1008,74(1) | 207 | 12070,3(1) | 1146,52(1) |
| 36 | 2050,8(1) | 1106,79(1) | 122 | 7089,8(1) | 987,03(1) | 208 | 12128,9(1) | 1145,67(1) |
| 37 | 2109,4(1) | 1121,41(1) | 123 | 7148,4(1) | 964,28(1) | 209 | 12187,5(1) | 1142,81(1) |
| 38 | 2168,0(1) | 1139,20(1) | 124 | 7207,0(1) | 943,80(1) | 210 | 12246,1(1) | 1139,18(1) |
| 39 | 2226,6(1) | 1158,59(1) | 125 | 7265,6(1) | 928,82(1) | 211 | 12304,7(1) | 1136,04(1) |
| 40 | 2285,2(1) | 1179,65(1) | 126 | 7324,2(1) | 913,99(1) | 212 | 12363,3(1) | 1128,33(1) |
| 41 | 2343,8(1) | 1193,71(1) | 127 | 7382,8(1) | 897,41(1) | 213 | 12421,9(1) | 1118,79(1) |
| 42 | 2402,3(1) | 1203,92(1) | 128 | 7441,4(1) | 877,87(1) | 214 | 12480,5(1) | 1108,47(1) |
| 43 | 2460,9(1) | 1215,11(1) | 129 | 7500,0(1) | 857,31(1) | 215 | 12539,1(1) | 1097,36(1) |
| 44 | 2519,5(1) | 1223,73(1) | 130 | 7558,6(1) | 836,59(1) | 216 | 12597,7(1) | 1085,47(1) |
| 45 | 2578,1(1) | 1236,12(1) | 131 | 7617,2(1) | 815,72(1) | 217 | 12656,3(1) | 1078,71(1) |
| 46 | 2636,7(1) | 1247,40(1) | 132 | 7675,8(1) | 794,31(1) | 218 | 12714,8(1) | 1072,26(1) |
| 47 | 2695,3(1) | 1255,48(1) | 133 | 7734,4(1) | 771,36(1) | 219 | 12773,4(1) | 1068,49(1) |
| 48 | 2753,9(1) | 1258,23(1) | 134 | 7793,0(1) | 746,68(1) | 220 | 12832,0(1) | 1063,66(1) |
| 49 | 2812,5(1) | 1256,24(1) | 135 | 7851,6(1) | 720,10(1) | 221 | 12890,6(1) | 1060,65(1) |
| 50 | 2871,1(1) | 1260,83(1) | 136 | 7910,2(1) | 692,69(1) | 222 | 12949,2(1) | 1058,10(1) |
| 51 | 2929,7(1) | 1265,58(1) | 137 | 7968,8(1) | 662,60(1) | 223 | 13007,8(1) | 1053,84(1) |
| 52 | 2988,3(1) | 1267,37(1) | 138 | 8027,3(1) | 633,63(1) | 224 | 13066,4(1) | 1052,63(1) |
| 53 | 3046,9(1) | 1265,89(1) | 139 | 8085,9(1) | 611,77(1) | 225 | 13125,0(1) | 1050,16(1) |
| 54 | 3105,5(1) | 1263,29(1) | 140 | 8144,5(1) | 600,57(1) | 226 | 13183,6(1) | 1046,19(1) |
| 55 | 3164,1(1) | 1263,21(1) | 141* | 8203,1(1) | 594,21(1) | 227 | 13242,2(1) | 1045,99(1) |
| 56 | 3222,7(1) | 1276,65(1) | 142 | 8261,7(1) | 597,93(1) | 228 | 13300,8(1) | 1044,42(1) |
| 57 | 3281,3(1) | 1282,21(1) | 143 | 8320,3(1) | 603,45(1) | 229 | 13359,4(1) | 1044,76(1) |
| 58 | 3339,8(1) | 1286,11(1) | 144 | 8378,9(1) | 605,49(1) | 230 | 13418,0(1) | 1043,71(1) |
| 59 | 3398,4(1) | 1285,14(1) | 145 | 8437,5(1) | 607,65(1) | 231 | 13476,6(1) | 1040,05(1) |
| 60 | 3457,0(1) | 1284,88(1) | 146 | 8496,1(1) | 611,74(1) | 232 | 13535,2(1) | 1039,41(1) |
| 61 | 3515,6(1) | 1281,27(1) | 147 | 8554,7(1) | 621,29(1) | 233 | 13593,8(1) | 1041,38(1) |
| 62 | 3574,2(1) | 1284,06(1) | 148 | 8613,3(1) | 628,68(1) | 234 | 13652,3(1) | 1044,01(1) |
| 63 | 3632,8(1) | 1283,31(1) | 149 | 8671,9(1) | 647,24(1) | 235 | 13710,9(1) | 1054,44(1) |
| 64 | 3691,4(1) | 1285,48(1) | 150 | 8730,5(1) | 663,68(1) | 236 | 13769,5(1) | 1066,67(1) |
| 65 | 3750,0(1) | 1287,94(1) | 151 | 8789,1(1) | 666,21(1) | 237 | 13828,1(1) | 1080,66(1) |
| 66 | 3808,6(1) | 1290,23(1) | 152 | 8847,7(1) | 663,40(1) | 238 | 13886,7(1) | 1090,50(1) |
| 67 | 3867,2(1) | 1289,68(1) | 153 | 8906,3(1) | 658,60(1) | 239 | 13945,3(1) | 1103,63(1) |
| 68 | 3925,8(1) | 1280,01(1) | 154 | 8964,8(1) | 650,19(1) | 240 | 14003,9(1) | 1113,79(1) |
| 69 | 3984,4(1) | 1266,21(1) | 155 | 9023,4(1) | 642,58(1) | 241 | 14062,5(1) | 1121,92(1) |
| 70 | 4043,0(1) | 1256,61(1) | 156 | 9082,0(1) | 633,66(1) | 242 | 14121,1(1) | 1129,62(1) |
| 71 | 4101,6(1) | 1252,20(1) | 157 | 9140,6(1) | 623,58(1) | 243 | 14179,7(1) | 1133,37(1) |
| 72 | 4160,2(1) | 1253,53(1) | 158 | 9199,2(1) | 614,88(1) | 244 | 14238,3(1) | 1135,27(1) |
| 73 | 4218,8(1) | 1252,14(1) | 159 | 9257,8(1) | 616,21(1) | 245 | 14296,9(1) | 1139,21(1) |
| 74 | 4277,3(1) | 1251,72(1) | 160 | 9316,4(1) | 627,36(1) | 246 | 14355,5(1) | 1139,28(1) |
| 75 | 4335,9(1) | 1255,04(1) | 161 | 9375,0(1) | 658,45(1) | 247 | 14414,1(1) | 1134,54(1) |
| 76 | 4394,5(1) | 1256,71(1) | 162 | 9433,6(1) | 703,26(1) | 248 | 14472,7(1) | 1132,10(1) |
| 77 | 4453,1(1) | 1256,47(1) | 163 | 9492,2(1) | 748,24(1) | 249 | 14531,3(1) | 1132,28(1) |
| 78 | 4511,7(1) | 1258,21(1) | 164 | 9550,8(1) | 774,78(1) | 250 | 14589,8(1) | 1136,87(1) |
| 79 | 4570,3(1) | 1261,56(1) | 165 | 9609,4(1) | 792,09(1) | 251 | 14648,4(1) | 1143,80(1) |
| 80 | 4628,9(1) | 1264,21(1) | 166 | 9668,0(1) | 805,58(1) | 252 | 14707,0(1) | 1148,25(1) |
| 81 | 4687,5(1) | 1273,54(1) | 167 | 9726,6(1) | 817,90(1) | 253 | 14765,6(1) | 1150,44(1) |
| 82 | 4746,1(1) | 1286,20(1) | 168 | 9785,2(1) | 828,44(1) | 254 | 14824,2(1) | 1153,24(1) |

| 83 | 4804,7(1) | 1294,37(1) | 169 | 9843,8(1) | 837,51(1) | 255 | 14882,8(1) | 1156,60(1) |
| 84 | 4863,3(1) | 1297,04(1) | 170 | 9902,3(1) | 850,43(1) | 256 | 14941,4(1) | 1162,83(1) |
| 85 | 4921,9(1) | 1298,20(1) | 171 | 9960,9(1) | 866,05(1) | | | |
| 86 | 4980,5(1) | 1303,87(1) | 172 | 10019,5(1) | 879,03(1) | | | |

\* Valor mínimo utilizado como referência para definir o início do degrau.

**Tabela 23.** Dados utilizados para plotar o gráfico da Figura 13 c.

| Medida | Posição x (nm) | Altura z (nm) | Medida | Posição x (nm) | Altura z (nm) | Medida | Posição x (nm) | Altura z (nm) |
|---|---|---|---|---|---|---|---|---|
| 1 | 0,0(1) | 847,36(1) | 71 | 4103,8(1) | 1095,30(1) | 141 | 8207,6(1) | 467,62(1) |
| 2 | 58,6(1) | 863,38(1) | 72 | 4162,4(1) | 1087,12(1) | 142 | 8266,2(1) | 486,94(1) |
| 3 | 117,3(1) | 872,79(1) | 73 | 4221,0(1) | 1081,74(1) | 143 | 8324,8(1) | 508,31(1) |
| 4 | 175,9(1) | 884,87(1) | 74 | 4279,7(1) | 1079,44(1) | 144 | 8383,4(1) | 531,84(1) |
| 5 | 234,5(1) | 895,11(1) | 75 | 4338,3(1) | 1071,31(1) | 145 | 8442,1(1) | 554,96(1) |
| 6 | 293,1(1) | 904,24(1) | 76 | 4396,9(1) | 1062,08(1) | 146 | 8500,7(1) | 580,57(1) |
| 7 | 351,8(1) | 912,15(1) | 77 | 4455,5(1) | 1052,90(1) | 147 | 8559,3(1) | 605,34(1) |
| 8 | 410,4(1) | 916,87(1) | 78 | 4514,2(1) | 1046,87(1) | 148 | 8618,0(1) | 627,96(1) |
| 9 | 469,0(1) | 921,46(1) | 79 | 4572,8(1) | 1044,97(1) | 149 | 8676,6(1) | 646,58(1) |
| 10 | 527,6(1) | 925,35(1) | 80 | 4631,4(1) | 1038,59(1) | 150 | 8735,2(1) | 664,56(1) |
| 11 | 586,3(1) | 925,01(1) | 81 | 4690,0(1) | 1034,35(1) | 151 | 8793,8(1) | 685,66(1) |
| 12 | 644,9(1) | 926,00(1) | 82 | 4748,7(1) | 1039,98(1) | 152 | 8852,5(1) | 702,78(1) |
| 13 | 703,5(1) | 929,65(1) | 83 | 4807,3(1) | 1039,79(1) | 153 | 8911,1(1) | 717,61(1) |
| 14 | 762,1(1) | 935,61(1) | 84 | 4865,9(1) | 1039,11(1) | 154 | 8969,7(1) | 731,58(1) |
| 15 | 820,8(1) | 946,11(1) | 85 | 4924,5(1) | 1047,16(1) | 155 | 9028,3(1) | 742,03(1) |
| 16 | 879,4(1) | 959,07(1) | 86 | 4983,2(1) | 1048,72(1) | 156 | 9087,0(1) | 752,88(1) |
| 17 | 938,0(1) | 972,59(1) | 87 | 5041,8(1) | 1048,48(1) | 157 | 9145,6(1) | 764,21(1) |
| 18 | 996,6(1) | 984,34(1) | 88 | 5100,4(1) | 1055,62(1) | 158 | 9204,2(1) | 775,51(1) |
| 19 | 1055,3(1) | 998,21(1) | 89 | 5159,0(1) | 1065,58(1) | 159 | 9262,8(1) | 789,65(1) |
| 20 | 1113,9(1) | 1005,43(1) | 90 | 5217,7(1) | 1070,18(1) | 160 | 9321,5(1) | 806,56(1) |
| 21 | 1172,5(1) | 1012,23(1) | 91 | 5276,3(1) | 1075,85(1) | 161 | 9380,1(1) | 821,74(1) |
| 22 | 1231,1(1) | 1018,03(1) | 92 | 5334,9(1) | 1085,03(1) | 162 | 9438,7(1) | 833,50(1) |
| 23 | 1289,8(1) | 1019,96(1) | 93 | 5393,5(1) | 1090,32(1) | 163 | 9497,3(1) | 851,33(1) |
| 24 | 1348,4(1) | 1021,26(1) | 94 | 5452,2(1) | 1090,73(1) | 164 | 9556,0(1) | 870,66(1) |
| 25 | 1407,0(1) | 1026,86(1) | 95 | 5510,8(1) | 1087,01(1) | 165 | 9614,6(1) | 888,81(1) |
| 26 | 1465,6(1) | 1036,36(1) | 96 | 5569,4(1) | 1078,68(1) | 166 | 9673,2(1) | 907,33(1) |
| 27 | 1524,3(1) | 1047,60(1) | 97 | 5628,1(1) | 1074,18(1) | 167 | 9731,8(1) | 923,57(1) |
| 28 | 1582,9(1) | 1053,10(1) | 98 | 5686,7(1) | 1073,48(1) | 168 | 9790,5(1) | 938,72(1) |
| 29 | 1641,5(1) | 1053,35(1) | 99 | 5745,3(1) | 1071,06(1) | 169 | 9849,1(1) | 955,14(1) |
| 30 | 1700,1(1) | 1053,88(1) | 100 | 5803,9(1) | 1067,02(1) | 170 | 9907,7(1) | 971,51(1) |
| 31 | 1758,8(1) | 1059,76(1) | 101 | 5862,6(1) | 1062,54(1) | 171 | 9966,3(1) | 987,88(1) |
| 32 | 1817,4(1) | 1064,69(1) | 102 | 5921,2(1) | 1058,01(1) | 172 | 10025,0(1) | 1004,61(1) |
| 33 | 1876,0(1) | 1069,87(1) | 103 | 5979,8(1) | 1052,05(1) | 173 | 10083,6(1) | 1021,48(1) |
| 34 | 1934,6(1) | 1074,44(1) | 104 | 6038,4(1) | 1043,22(1) | 174 | 10142,2(1) | 1038,90(1) |
| 35 | 1993,3(1) | 1072,09(1) | 105 | 6097,1(1) | 1040,33(1) | 175 | 10200,8(1) | 1049,39(1) |
| 36 | 2051,9(1) | 1076,63(1) | 106 | 6155,7(1) | 1033,54(1) | 176 | 10259,5(1) | 1057,34(1) |
| 37 | 2110,5(1) | 1084,02(1) | 107 | 6214,3(1) | 1015,47(1) | 177 | 10318,1(1) | 1068,78(1) |
| 38 | 2169,1(1) | 1090,94(1) | 108 | 6272,9(1) | 987,82(1) | 178 | 10376,7(1) | 1081,98(1) |
| 39 | 2227,8(1) | 1100,86(1) | 109 | 6331,6(1) | 959,96(1) | 179 | 10435,3(1) | 1090,33(1) |
| 40 | 2286,4(1) | 1106,26(1) | 110 | 6390,2(1) | 932,90(1) | 180 | 10494,0(1) | 1099,21(1) |
| 41 | 2345,0(1) | 1110,32(1) | 111 | 6448,8(1) | 902,79(1) | 181 | 10552,6(1) | 1115,18(1) |
| 42 | 2403,6(1) | 1113,12(1) | 112 | 6507,4(1) | 868,18(1) | 182 | 10611,2(1) | 1134,53(1) |
| 43 | 2462,3(1) | 1112,54(1) | 113 | 6566,1(1) | 817,57(1) | 183 | 10669,8(1) | 1151,89(1) |
| 44 | 2520,9(1) | 1114,72(1) | 114 | 6624,7(1) | 753,55(1) | 184 | 10728,5(1) | 1161,84(1) |
| 45 | 2579,5(1) | 1120,51(1) | 115 | 6683,3(1) | 682,50(1) | 185 | 10787,1(1) | 1170,10(1) |
| 46 | 2638,1(1) | 1128,96(1) | 116 | 6741,9(1) | 599,24(1) | 186 | 10845,7(1) | 1180,17(1) |
| 47 | 2696,8(1) | 1134,25(1) | 117 | 6800,6(1) | 525,58(1) | 187 | 10904,3(1) | 1192,41(1) |
| 48 | 2755,4(1) | 1140,51(1) | 118 | 6859,2(1) | 454,14(1) | 188 | 10963,0(1) | 1205,48(1) |
| 49 | 2814,0(1) | 1146,78(1) | 119 | 6917,8(1) | 365,15(1) | 189 | 11021,6(1) | 1216,51(1) |
| 50 | 2872,7(1) | 1151,71(1) | 120 | 6976,4(1) | 279,18(1) | 190 | 11080,2(1) | 1222,79(1) |
| 51 | 2931,3(1) | 1158,78(1) | 121 | 7035,1(1) | 227,96(1) | 191 | 11138,8(1) | 1224,73(1) |
| 52 | 2989,9(1) | 1166,04(1) | 122 | 7093,7(1) | 198,37(1) | 192 | 11197,5(1) | 1225,37(1) |

| | | | | | | | |
|---|---|---|---|---|---|---|---|
| 53 | 3048,5(1) | 1168,64(1) | 123 | 7152,3(1) | 181,20(1) | 193 | 11256,1(1) | 1228,63(1) |
| 54 | 3107,2(1) | 1169,50(1) | 124* | 7210,9(1) | 174,33(1) | 194 | 11314,7(1) | 1231,06(1) |
| 55 | 3165,8(1) | 1170,79(1) | 125 | 7269,6(1) | 180,67(1) | 195 | 11373,4(1) | 1232,39(1) |
| 56 | 3224,4(1) | 1171,73(1) | 126 | 7328,2(1) | 196,06(1) | 196 | 11432,0(1) | 1231,49(1) |
| 57 | 3283,0(1) | 1170,03(1) | 127 | 7386,8(1) | 209,69(1) | 197 | 11490,6(1) | 1225,37(1) |
| 58 | 3341,7(1) | 1169,24(1) | 128 | 7445,4(1) | 232,42(1) | 198 | 11549,2(1) | 1219,06(1) |
| 59 | 3400,3(1) | 1167,74(1) | 129 | 7504,1(1) | 257,86(1) | 199 | 11607,9(1) | 1212,30(1) |
| 60 | 3458,9(1) | 1166,64(1) | 130 | 7562,7(1) | 280,32(1) | 200 | 11666,5(1) | 1199,17(1) |
| 61 | 3517,5(1) | 1165,85(1) | 131 | 7621,3(1) | 305,61(1) | 201 | 11725,1(1) | 1179,99(1) |
| 62 | 3576,2(1) | 1162,88(1) | 132 | 7679,9(1) | 324,54(1) | 202 | 11783,7(1) | 1163,54(1) |
| 63 | 3634,8(1) | 1158,41(1) | 133 | 7738,6(1) | 341,93(1) | 203 | 11842,4(1) | 1145,47(1) |
| 64 | 3693,4(1) | 1150,60(1) | 134 | 7797,2(1) | 357,64(1) | 204 | 11901,0(1) | 1122,34(1) |
| 65 | 3752,0(1) | 1136,73(1) | 135 | 7855,8(1) | 372,52(1) | 205 | 11959,6(1) | 1098,83(1) |
| 66 | 3810,7(1) | 1127,74(1) | 136 | 7914,4(1) | 388,02(1) | 206 | 12018,2(1) | 1072,66(1) |
| 67 | 3869,3(1) | 1123,38(1) | 137 | 7973,1(1) | 402,19(1) | 207 | 12076,9(1) | 1040,12(1) |
| 68 | 3927,9(1) | 1118,26(1) | 138 | 8031,7(1) | 415,09(1) | 208 | 12135,5(1) | 1002,95(1) |
| 69 | 3986,5(1) | 1110,29(1) | 139 | 8090,3(1) | 429,32(1) | 209 | 12194,1(1) | 964,18(1) |
| 70 | 4045,2(1) | 1101,17(1) | 140 | 8148,9(1) | 446,89(1) | 141 | 8207,6(1) | 467,62(1) |

* Valor mínimo utilizado como referência para definir o início do degrau.

**Tabela 24.** Dados utilizados para plotar o gráfico da Figura 12 c.

| Medida | Posição x (nm) | Altura z (nm) | Medida | Posição x (nm) | Altura z (nm) | Medida | Posição x (nm) | Altura z (nm) |
|---|---|---|---|---|---|---|---|---|
| 1 | 0,0(1) | 706,77(1) | 79 | 4569,0(1) | 409,95(1) | 157 | 9138,0(1) | 551,11(1) |
| 2 | 58,6(1) | 700,13(1) | 80 | 4627,6(1) | 412,75(1) | 158 | 9196,6(1) | 553,32(1) |
| 3 | 117,2(1) | 692,05(1) | 81 | 4686,1(1) | 413,34(1) | 159 | 9255,1(1) | 556,48(1) |
| 4 | 175,7(1) | 682,19(1) | 82 | 4744,7(1) | 417,00(1) | 160 | 9313,7(1) | 559,94(1) |
| 5 | 234,3(1) | 670,77(1) | 83 | 4803,3(1) | 423,13(1) | 161 | 9372,3(1) | 562,99(1) |
| 6 | 292,9(1) | 660,24(1) | 84 | 4861,9(1) | 429,66(1) | 162 | 9430,9(1) | 567,04(1) |
| 7 | 351,5(1) | 648,44(1) | 85 | 4920,5(1) | 436,98(1) | 163 | 9489,5(1) | 570,30(1) |
| 8 | 410,0(1) | 636,05(1) | 86 | 4979,0(1) | 445,04(1) | 164 | 9548,0(1) | 570,35(1) |
| 9 | 468,6(1) | 621,96(1) | 87 | 5037,6(1) | 450,01(1) | 165 | 9606,6(1) | 573,56(1) |
| 10 | 527,2(1) | 603,41(1) | 88 | 5096,2(1) | 453,44(1) | 166 | 9665,2(1) | 572,85(1) |
| 11 | 585,8(1) | 586,26(1) | 89 | 5154,8(1) | 456,60(1) | 167 | 9723,8(1) | 570,80(1) |
| 12 | 644,3(1) | 572,15(1) | 90 | 5213,3(1) | 457,75(1) | 168 | 9782,3(1) | 572,65(1) |
| 13 | 702,9(1) | 557,55(1) | 91 | 5271,9(1) | 457,86(1) | 169 | 9840,9(1) | 577,39(1) |
| 14 | 761,5(1) | 539,46(1) | 92 | 5330,5(1) | 456,81(1) | 170 | 9899,5(1) | 582,60(1) |
| 15 | 820,1(1) | 524,88(1) | 93 | 5389,1(1) | 456,28(1) | 171 | 9958,1(1) | 584,88(1) |
| 16 | 878,7(1) | 512,98(1) | 94 | 5447,6(1) | 459,19(1) | 172 | 10016,6(1) | 586,91(1) |
| 17 | 937,2(1) | 498,70(1) | 95 | 5506,2(1) | 464,42(1) | 173 | 10075,2(1) | 586,23(1) |
| 18 | 995,8(1) | 473,38(1) | 96 | 5564,8(1) | 469,94(1) | 174 | 10133,8(1) | 583,40(1) |
| 19 | 1054,4(1) | 432,41(1) | 97 | 5623,4(1) | 473,18(1) | 175 | 10192,4(1) | 577,69(1) |
| 20 | 1113,0(1) | 382,38(1) | 98 | 5682,0(1) | 476,95(1) | 176 | 10251,0(1) | 570,89(1) |
| 21 | 1171,5(1) | 334,70(1) | 99 | 5740,5(1) | 478,02(1) | 177 | 10309,5(1) | 564,42(1) |
| 22 | 1230,1(1) | 293,81(1) | 100 | 5799,1(1) | 475,91(1) | 178 | 10368,1(1) | 558,32(1) |
| 23 | 1288,7(1) | 259,63(1) | 101 | 5857,7(1) | 473,33(1) | 179 | 10426,7(1) | 549,28(1) |
| 24 | 1347,3(1) | 234,82(1) | 102 | 5916,3(1) | 473,97(1) | 180 | 10485,3(1) | 539,58(1) |
| 25 | 1405,8(1) | 220,73(1) | 103 | 5974,8(1) | 484,21(1) | 181 | 10543,8(1) | 532,42(1) |
| 26 | 1464,4(1) | 216,15(1) | 104 | 6033,4(1) | 497,22(1) | 182 | 10602,4(1) | 528,93(1) |
| 27 | 1523,0(1) | 216,43(1) | 105 | 6092,0(1) | 510,30(1) | 183 | 10661,0(1) | 524,30(1) |
| 28 | 1581,6(1) | 218,28(1) | 106 | 6150,6(1) | 528,73(1) | 184 | 10719,6(1) | 518,99(1) |
| 29 | 1640,2(1) | 219,07(1) | 107 | 6209,1(1) | 548,53(1) | 185 | 10778,1(1) | 514,51(1) |
| 30 | 1698,7(1) | 219,81(1) | 108 | 6267,7(1) | 565,80(1) | 186 | 10836,7(1) | 512,78(1) |
| 31 | 1757,3(1) | 219,69(1) | 109 | 6326,3(1) | 587,23(1) | 187 | 10895,3(1) | 514,12(1) |
| 32 | 1815,9(1) | 218,48(1) | 110 | 6384,9(1) | 611,07(1) | 188 | 10953,9(1) | 517,12(1) |
| 33 | 1874,5(1) | 214,51(1) | 111 | 6443,5(1) | 633,38(1) | 189 | 11012,5(1) | 523,66(1) |
| 34 | 1933,0(1) | 208,23(1) | 112 | 6502,0(1) | 652,70(1) | 190 | 11071,0(1) | 526,02(1) |
| 35 | 1991,6(1) | 201,40(1) | 113 | 6560,6(1) | 667,50(1) | 191 | 11129,6(1) | 525,95(1) |
| 36 | 2050,2(1) | 194,72(1) | 114 | 6619,2(1) | 679,97(1) | 192 | 11188,2(1) | 526,04(1) |
| 37 | 2108,8(1) | 189,23(1) | 115 | 6677,8(1) | 694,29(1) | 193 | 11246,8(1) | 527,80(1) |
| 38 | 2167,3(1) | 189,90(1) | 116 | 6736,3(1) | 703,52(1) | 194 | 11305,3(1) | 528,04(1) |
| 39 | 2225,9(1) | 192,54(1) | 117 | 6794,9(1) | 711,84(1) | 195 | 11363,9(1) | 527,45(1) |
| 40 | 2284,5(1) | 193,77(1) | 118 | 6853,5(1) | 721,62(1) | 196 | 11422,5(1) | 528,82(1) |
| 41 | 2343,1(1) | 193,19(1) | 119 | 6912,1(1) | 728,19(1) | 197 | 11481,1(1) | 531,03(1) |
| 42 | 2401,7(1) | 191,01(1) | 120 | 6970,6(1) | 736,45(1) | 198 | 11539,6(1) | 531,09(1) |
| 43 | 2460,2(1) | 188,04(1) | 121 | 7029,2(1) | 740,97(1) | 199 | 11598,2(1) | 530,36(1) |
| 44 | 2518,8(1) | 184,04(1) | 122 | 7087,8(1) | 741,47(1) | 200 | 11656,8(1) | 528,50(1) |
| 45 | 2577,4(1) | 179,05(1) | 123 | 7146,4(1) | 737,94(1) | 201 | 11715,4(1) | 525,02(1) |
| 46 | 2636,0(1) | 174,83(1) | 124 | 7205,0(1) | 735,67(1) | 202 | 11774,0(1) | 519,38(1) |
| 47 | 2694,5(1) | 170,64(1) | 125 | 7263,5(1) | 732,93(1) | 203 | 11832,5(1) | 510,16(1) |
| 48 | 2753,1(1) | 169,30(1) | 126 | 7322,1(1) | 728,71(1) | 204 | 11891,1(1) | 500,48(1) |
| 49 | 2811,7(1) | 168,27(1) | 127 | 7380,7(1) | 721,27(1) | 205 | 11949,7(1) | 494,39(1) |
| 50 | 2870,3(1) | 162,71(1) | 128 | 7439,3(1) | 711,39(1) | 206 | 12008,3(1) | 486,71(1) |
| 51 | 2928,8(1) | 154,77(1) | 129 | 7497,8(1) | 701,64(1) | 207 | 12066,8(1) | 476,71(1) |
| 52* | 2987,4(1) | 152,60(1) | 130 | 7556,4(1) | 692,04(1) | 208 | 12125,4(1) | 463,23(1) |
| 53 | 3046,0(1) | 161,54(1) | 131 | 7615,0(1) | 682,27(1) | 209 | 12184,0(1) | 451,19(1) |
| 54 | 3104,6(1) | 180,44(1) | 132 | 7673,6(1) | 671,02(1) | 210 | 12242,6(1) | 447,06(1) |
| 55 | 3163,2(1) | 200,68(1) | 133 | 7732,1(1) | 660,05(1) | 211 | 12301,1(1) | 445,68(1) |
| 56 | 3221,7(1) | 218,94(1) | 134 | 7790,7(1) | 648,08(1) | 212 | 12359,7(1) | 442,83(1) |
| 57 | 3280,3(1) | 233,30(1) | 135 | 7849,3(1) | 640,38(1) | 213 | 12418,3(1) | 439,56(1) |
| 58 | 3338,9(1) | 245,35(1) | 136 | 7907,9(1) | 635,97(1) | 214 | 12476,9(1) | 438,23(1) |
| 59 | 3397,5(1) | 265,51(1) | 137 | 7966,5(1) | 630,30(1) | 215 | 12535,5(1) | 439,72(1) |
| 60 | 3456,0(1) | 292,56(1) | 138 | 8025,0(1) | 624,02(1) | 216 | 12594,0(1) | 441,48(1) |
| 61 | 3514,6(1) | 318,58(1) | 139 | 8083,6(1) | 618,59(1) | 217 | 12652,6(1) | 444,21(1) |
| 62 | 3573,2(1) | 344,96(1) | 140 | 8142,2(1) | 611,80(1) | 218 | 12711,2(1) | 448,75(1) |
| 63 | 3631,8(1) | 368,29(1) | 141 | 8200,8(1) | 607,56(1) | 219 | 12769,8(1) | 456,49(1) |
| 64 | 3690,3(1) | 387,51(1) | 142 | 8259,3(1) | 602,12(1) | 220 | 12828,3(1) | 466,23(1) |
| 65 | 3748,9(1) | 404,05(1) | 143 | 8317,9(1) | 597,58(1) | 221 | 12886,9(1) | 473,94(1) |
| 66 | 3807,5(1) | 414,48(1) | 144 | 8376,5(1) | 592,79(1) | 222 | 12945,5(1) | 480,28(1) |
| 67 | 3866,1(1) | 420,40(1) | 145 | 8435,1(1) | 585,70(1) | 223 | 13004,1(1) | 483,61(1) |
| 68 | 3924,7(1) | 424,65(1) | 146 | 8493,6(1) | 579,07(1) | 224 | 13062,6(1) | 482,45(1) |
| 69 | 3983,2(1) | 428,49(1) | 147 | 8552,2(1) | 573,59(1) | 225 | 13121,2(1) | 483,35(1) |
| 70 | 4041,8(1) | 432,99(1) | 148 | 8610,8(1) | 570,15(1) | 226 | 13179,8(1) | 484,16(1) |
| 71 | 4100,4(1) | 435,30(1) | 149 | 8669,4(1) | 565,52(1) | 227 | 13238,4(1) | 481,27(1) |
| 72 | 4159,0(1) | 435,18(1) | 150 | 8728,0(1) | 562,39(1) | 228 | 13297,0(1) | 480,21(1) |
| 73 | 4217,5(1) | 432,05(1) | 151 | 8786,5(1) | 563,25(1) | 229 | 13355,5(1) | 480,12(1) |
| 74 | 4276,1(1) | 426,16(1) | 152 | 8845,1(1) | 562,29(1) | 230 | 13414,1(1) | 481,40(1) |
| 75 | 4334,7(1) | 417,26(1) | 153 | 8903,7(1) | 559,94(1) | 231 | 13472,7(1) | 479,63(1) |
| 76 | 4393,3(1) | 410,03(1) | 154 | 8962,3(1) | 557,19(1) | 232 | 13531,3(1) | 476,74(1) |
| 77 | 4451,8(1) | 405,53(1) | 155 | 9020,8(1) | 554,70(1) | 233 | 13589,8(1) | 474,17(1) |
| 78 | 4510,4(1) | 406,27(1) | 156 | 9079,4(1) | 551,99(1) | | | |

\* Valor mínimo utilizado como referência para definir o início do degrau.

Para a propagação de erro dos cálculos, considerou-se as incertezas do microscópio Shimadzu SPM-9700HT como sendo 0,2 nm nas coordenadas "x" e "y", e de 0,01 nm na coordenada "z", conforme informações da fabricante (40). A partir dos perfis altimétricos, definiu-se o ponto de menor valor em "z" (destacado com " * " nas tabelas) como o ponto mínimo de referência para o cálculo do degrau , sendo possível obter a altura mínima relativa de referência ($h_b$) e sua incerteza instrumental ( $\sigma_{h_b}$) de 0,01 nm. A partir dele, determinou-se qual região do perfil seria considerada para a análise da deposição de grafite, destacada nas Figuras Figura *12*, Figura **13** e Figura **14** c. Com os dados contidos nessa região, calculou-se a média aritmética da altura relativa do degrau ($\overline{h_d}$), bem como o seu desvio padrão da média $\sigma_{\overline{h_d}}$ (Equação ( 20). Para calcular a incerteza total ($u_{total}$) a associada a $\overline{h_d}$ , levou-se em consideração tanto o $\sigma_{\overline{h_d}}$ quanto a incerteza do instrumento ($u_{ins} = 0,01$ nm), conforme descrito na Equação ( 19. Posteriormente, realizou-se a subtração $\overline{h_d} - h_b$, afim de obter-se o valor absoluto da altura do degrau. Desta forma, foi necessário realizar a propagação do da operação de subtração, através da seguinte equação:

$$u_{\overline{h_d}-h_b} = \sqrt{u_{\overline{h_d}}^2 + u_{h_b}^2} \qquad (29)$$

Em que $u_{\overline{h_d}-h_b}$ é a incerteza do valor resultante da operação de subtração, $u_{\overline{h_d}}$ é a incerteza total de $\overline{h_d}$ e $u_{h_b}$ é a incerteza instrumental de $h_b$. Os valores obtidos para cada grandeza envolvida estão apresentados na **Tabela 25**.

**Tabela 25.** Grandezas obtidas a partir dos perfis altimétricos.

| Figura | Grandeza | Valor | Unidade de Medida | Observação |
|---|---|---|---|---|
| | n | 140 | | Quantidade de dados |
| | $h_b$ | 594,2 | nm | |
| | $\sigma_{h_b}$ | 0,01 | nm | Incerteza instrumental |
| | $\overline{h_d}$ | 1094,20573 | nm | Média aritmética |
| | | 1094 | nm | Arredondamento |
| | $\sigma_{\overline{h_d}}$ | 17,41048 | nn | Equação ( 20 ) |
| | $\overline{h_d} \rightarrow u_{ins}$ | 0,01 | nm | |
| 12 | $\overline{h_d} \rightarrow u_{total}$ | 17,41048 | nm | Equação ( 19 ) |
| | | 17 | nm | Arredondamento |
| | $\overline{h_d} - h_b$ | 499,99373 | nm | |
| | | 500 | nm | Arredondamento |
| | $u_{\overline{h_d}-h_b}$ | 17,00000 | nm | Equação ( 29 |
| | | 17 | nm | Arredondamento |
| | $h$ | 500(17) | nm | Intervalo das medidas da Tabela 9 usado: 1 – 50 |
| | n | 124 | | Quantidade de dados |
| | $h_b$ | 174,3 | nm | |
| | $\sigma_{h_b}$ | 0,01 | nm | Incerteza instrumental |
| | $\overline{h_d}$ | 993,01487 | nm | Média aritmética |
| | | 993 | nm | Arredondamento |
| | $\sigma_{\overline{h_d}}$ | 18,86568 | nn | Equação ( 20 ) |
| | $\overline{h_d} \rightarrow u_{ins}$ | 0,01 | nm | |
| 13 | $\overline{h_d} \rightarrow u_{total}$ | 18,86568 | nm | Equação ( 19 ) |
| | | 19 | nm | Arredondamento |
| | $\overline{h_d} - h_b$ | 818,68987 | nm | |
| | | 819 | nm | Arredondamento |
| | $u_{\overline{h_d}-h_b}$ | 19,00000 | nm | Equação ( 29 |
| | | 19 | nm | Arredondamento |
| | $h$ | 819(19) | nm | Intervalo das medidas da Tabela 9 usado: 1 – 50 |

| | Grandeza | Valor | Unidade | Observação |
|---|---|---|---|---|
| | n | 182 | | Quantidade de dados |
| | $h_b$ | 152,60100 | nm | |
| | $\sigma_{h_b}$ | 0,01 | nm | Incerteza instrumental |
| | $\overline{h_d}$ | 516,03969 | nm | Média aritmética |
| | | 516 | nm | Arredondamento |
| | $\sigma_{\overline{h_d}}$ | 8,40611 | nn | Equação ( 20 ) |
| | $\overline{h_d} \to u_{ins}$ | 0,01 | nm | |
| 14 | $\overline{h_d} \to u_{total}$ | 8,40001 | nm | Equação ( 19 ) |
| | | 8 | nm | Arredondamento |
| | $\overline{h_d} - h_b$ | 363,43869 | nm | |
| | | 363 | nm | Arredondamento |
| | $u_{\overline{h_d} - h_b}$ | 8,00001 | nm | Equação ( 29 |
| | | 8 | nm | Arredondamento |
| | h | 363(8) | nm | Intervalo das medidas da Tabela 9 usado: 1 – 50 |

A partir dos dados da altura das camadas de grafite depositados sobre o papel vegetal e suas respectivas incertezas propagadas foi possível obter o valor médio da altura da camada de grafite por média ponderada pela incerteza (Equação ( 27), já que os valores obtidos apresentaram incertezas distintas e calcular a incerteza total através da Equação ( 28. Os valores estatísticos obtidos são apresentados na Tabela 26.

**Tabela 26.** Grandezas estatísitcas dos cálculos da média de alturas obtidas a partir das medidas dos perfis de altimetria das Figuras Figura **12**, Figura *13* e Figura *14*.

| Grandeza | Valor (nm) | Observação |
|---|---|---|
| h | 500(17) | Figura 12 |
| h | 819(19) | Figura 13 |
| h | 363(8) | Figura 14 |
| $\overline{h}$ | 442,48676 | Equação ( 27 ) |
| | 440 | Arredondamento |
| $\sigma_{\bar{x}}$ | 45,75551 | Equação ( 28 ) |
| | 40 | Arredondamento |
| h | 440(40) | Valor médio da altura da camada de grafite |